\documentclass[%
twocolumn,
superscriptaddress,
nofootinbib,
 amsmath,amssymb,
 aps, prd
]{revtex4-2}

\usepackage{graphicx}
\usepackage{tensor}
\usepackage{siunitx}
\usepackage{xcolor} % Remove after notes are no longer needed
\usepackage[normalem]{ulem} % Remove after notes are no longer needed
\usepackage{lipsum} % Remove after filler text is no longer needed

\usepackage[pdfborder={0 0 0}, plainpages=false, hyperfigures=true]{hyperref}

\newcommand{\beq}{\begin{equation}}
\newcommand{\eeq}{\end{equation}}
\newcommand{\beqn}{\begin{eqnarray}}
\newcommand{\eeqn}{\end{eqnarray}}

\newcommand{\lo}{\mathrel{\raise.3ex\hbox{$<$}\mkern-14mu
    \lower0.6ex\hbox{$\sim$}}}
\newcommand{\go}{\mathrel{\raise.3ex\hbox{$>$}\mkern-14mu
    \lower0.6ex\hbox{$\sim$}}}

\newcommand{\WSU}{\affiliation{Department of Physics \& Astronomy,
	Washington State University, Pullman, Washington 99164, USA}}
\newcommand{\UNH}{\affiliation{Department of Physics \& Astronomy, University of New Hampshire, 9 Library Way, Durham NH 03824, USA}}
\newcommand{\TAPIR}{\affiliation{TAPIR, Walter Burke Institute for Theoretical Physics, MC 350-17, California Institute of Technology, Pasadena, California 91125, USA}}
\newcommand{\Cornell}{\affiliation{Cornell Center for Astrophysics and Planetary Science, Cornell University, Ithaca, New York, 14853, USA}}
\newcommand{\MPI}{\affiliation{Max Planck Institute for Gravitational Physics (Albert Einstein Institute), Am M{\H u}hlenberg 1, 14476 Potsdam, Germany}}

\newcommand{\UCBP}{\affiliation{Department of Physics, University of California, Berkeley, Berkeley, CA 94720, USA}}

\begin{document}
\title{Impact of neutrino-electron scattering and an improved treatment of pair processes on binary neutron star mergers}
\author{Francois Foucart}\UNH
\author{Samantha Rath}\UNH
\author{Rowan Davidson}\UNH
\author{Patrick Chi-Kit Cheong}\UNH\UCBP
\author{Matthew D. Duez}\WSU
\author{Lawrence Kidder}\Cornell
\author{Harald Pfeiffer}\MPI
\author{Mark Scheel}\TAPIR

\begin{abstract}
Multimessenger observations of neutron star mergers are unique opportunities to constrain the properties of dense matter and the production site of heavy nuclei. To leverage these observations, we require reliable models of the electromagnetic signals powered by mergers. An important limitation to our ability to develop such models is the use of approximate neutrino physics in simulations. Here, we present simulations using an improved version of our Monte Carlo transport algorithm specifically designed to allow for more advanced on-the-fly calculations of reaction rates that use the simulated energy distribution of neutrinos, including in blocking factors, while still relying on approximations for the angular distribution of neutrinos. We use these new methods to include in simulations inelastic scattering of neutrinos on electrons, and to improve  our treatment of neutrino-antineutrino pair annihilation. We find that, without increasing the cost of simulations, we can marginally get to the point when the addition of a single packet represents a change $\Delta f_\nu<1$ in the angle-integrated distribution function, at the cost of increased shot noise in the coupling to the fluid. With inelastic scattering and a better treatment of pair processes, we find a reduction in the average energy and total luminosity of heavy-lepton neutrinos, and an increase in the amount of mass ejected -- here by $50\%$, although on a relatively low amount of total ejected mass $<0.005M_\odot$. In a separate set of simulations varying the total mass of the binary away from its prompt collapse threshold, we find rapid variations in the amount of ejected matter and in the geometry and composition of the outflows with the total mass of the system. Finally, we use the simulations with our more advanced transport scheme to study in more detail the energy spectrum of neutrinos across the merger remnant.
\end{abstract}

\maketitle

\section{Introduction}

The observation of neutron star mergers through gravitational wave and electromagnetic signals provides us with remarkable opportunities to study the laws of physics in extremely dense, neutron-rich environments. Neutron star mergers are also one of the leading candidates for the production site of about half of the heavy elements -- those produced through rapid neutron capture nucleosynthesis. The first multi-messenger observation of a neutron star merger demonstrated their potential to constrain the properties of dense matter~\cite{GW170817-NSRadius,2017ApJL2041}. The optical/infrared kilonova that followed the merger also confirmed that at least some r-process elements are produced in these mergers~\cite{2017Natur.551...80K}. While no other multi-messenger observation has materialized itself so far, kilonovae likely associated with neutron star mergers have been observed in the afterglow of gamma-ray bursts (including some long bursts, for which the physical progenitor is then very uncertain)~\cite{Rastinejad:2024zuk}. Additionally, one high-mass binary neutron star merger was observed solely in gravitational waves~\cite{Abbott:2020uma}.

To reliably extract information about dense matter and nucleosynthesis from kilonovae, we require robust models of the matter ejected during and after merger. More specifically, kilonova signals are very sensitive to the neutron-richness, geometry, and temperature of the outflows~\cite{2013ApJ...775...18B}. The accuracy of current models is however limited by both the input nuclear physics (e.g. properties of neutron-rich heavy nuclei) and the accuracy of neutron star merger simulations. For the latter, the main current roadblocks include the modeling of the growth and saturation of magnetic fields and the treatment of neutrinos. Reliable evolution of magnetic fields indeed require very high spatial resolution~\cite{Kiuchi:2023obe,Kiuchi:2026pgb}, while the evolution of neutrinos require solving the high-dimensional Boltzmann equation of radiation transport as well as capturing a broad range of neutrino-matter and neutrino-neutrino interactions that can be difficult to include in numerical simulations~\cite{Foucart:2022bth}.

Most state-of-the art numerical codes evolve neutrinos using an energy-integrated (`gray') two-moment scheme, in which the energy density, momentum density, and often number density of neutrinos are evolved~\cite{1981MNRAS.194..439T,shibata:11,FoucartM1:2015,Foucart:2016rxm,Radice:2021jtw}. Moment schemes capture the leading order effects of neutrinos on the merger and post-merger remnants, and on the matter outflows. Nevertheless, they rely on approximate analytical closures for the pressure tensor of neutrinos and, maybe more importantly, the energy spectrum of neutrinos. The latter is crucial to the accurate calculation of neutrino-matter and neutrino-neutrino interaction rates. Some reactions (e.g. $\nu+\bar\nu \rightarrow e^++e^-$) also depend sensitively on the angular distribution of neutrinos, which moment schemes can only partially capture. A more recent development has been the use of Monte Carlo methods for neutrino transport~\cite{Foucart:2022kon}. Monte Carlo transport can more easily remain energy-dependent, and converges to the true solution of Boltzmann's equation as the number of `packets' used to sample the neutrino distribution function increases. They have however a low convergence rate, especially in the presence of optically thick regions, making it costly to reach high accuracy. Current global simulations of neutron star mergers have been limited to low-accuracy Monte Carlo transport~\cite{Foucart:2022kon,Foucart:2022kon}. The situation is better for black hole-disk simulations~\cite{Miller:2019dpt}, due to the lack of optically thick regions, and for axisymmetric simulations~\cite{Kawaguchi:2024naa}.

In~\cite{Foucart:2025nub}, we argued that capturing the distribution function of neutrinos within a given grid cell of a simulation without radically increasing the cost of simulations was only possible if (a) we ignored the angular dependence of the distribution function and (b) the weighting of the `packets' was significantly modified, with a likely trade-off being increased shot noise in the coupling of neutrinos to the matter. Despite the difficulty of capturing the distribution function of neutrinos, doing so comes with some major advantages: the ability to compute interaction rates for reactions that have been so far ignored in simulations, or treated very approximately. Specifically, calculating interaction rates for inelastic scattering and $\nu\bar\nu$ pair production/annihilation requires knowledge of the neutrino distribution function. In~\cite{Cheong:2024cnb}, we demonstrated that for an isolated neutron star, inelastic scattering could have a significant effect of the production of neutrino-driven winds. In~\cite{Rath:2026vfr}, we analyzed snapshots of neutron star merger simulations, and showed that both inelastic scattering and a better treatment of pair processes would impact the evolution of heavy lepton neutrinos. The inclusion of these reactions is thus an important step in assessing the importance of different approximations made in merger simulations.

In this manuscript, we present the first merger simulation to include inelastic scattering as well as a treatment of  $\nu+\bar\nu \leftrightarrow e^++e^-$ that accounts for the energy distribution of neutrinos and antineutrinos. We do so approximately: for inelastic scattering, we assume isotropic distributions of neutrinos, relying on the fact that inelastic scattering is always subdominant with respect to elastic scattering in the conditions encountered in neutron star mergers. For pair annihilation, the more important dependence of the reaction rate on the relative direction of propagation of the neutrinos is only included through a local `form factor' that models the average correction from isotropic rates for all neutrinos within a spatial grid cell. Both of these changes are nevertheless major improvements on our ability to model neutrino physics, made possible by changes to the weighting of Monte Carlo packets in our simulations. 

As part of this project, we also performed a series of simulations of neutron star binaries with slightly varying total mass, but constant mass ratio and equation of state. These simulations were meant to go from the rapidly collapsing systems studied in~\cite{Foucart:2024npn} to a post-merger remnant with a $O(10\,{\rm ms})$ collapse time. This manuscript is organized as follow. In Sec.~\ref{sec:methods}, we present the changes made to our Monte Carlo code to handle our new set of reactions. In Sec.~\ref{sec:sim}, we briefly review our other numerical methods and the initial conditions evolved in this manuscript. In Sec.~\ref{sec:results}, we first discuss the impact of a changing total mass on the evolution of the post-merger remnant and on the  generation of outflows, then focus more on the impact of improved neutrino physics in the simulations. 

\section{Monte Carlo transport improvements}
\label{sec:methods}

The more advanced simulations presented here include a few important modifications to the Monte Carlo methods used in previous SpEC simulations and described in~\cite{Foucart:2021mcb}: an updated treatment of the coupling between neutrinos and matter, a modification of the weights assigned to each Monte Carlo packet, the approximate inclusion of inelastic scattering of neutrinos on electrons, and an improved treatment of the production and annihilation of neutrino-antineutrino pairs. We discuss each in more details in the following sections.

\subsection{Neutrino coupling to the fluid}

The simplest change is the coupling between neutrino packets and the fluid sector. This coupling can generally be written as
\beq
\nabla_\mu T^{\mu\nu}_{\rm fluid} = G^\nu;\,\,\nabla_\mu T^{\mu\nu}_{\rm rad} = - G^\nu; \nabla_\mu (\rho Y_e u^\mu) = C^0.
\eeq
with $T^{\mu\nu}_{\rm fluid}, T^{\mu\nu}_{\rm rad}$ the stress-energy tensors of the fluid and radiation sector, $G^\nu$ the rate of 4-momentum transfer between neutrinos and matter, and $C^0$ a scalar field capturing composition changes due to charged current reactions. In previous SpEC simulations, $G^\nu$ and $C^0$ were always calculated based on their {\it expectation value} for absorption and scattering events, rather than explicitly accounting for the absortion and scattering events experienced by a given Monte Carlo packet. This was done to avoid excessive shot noise in the evolution of the composition and temperature of the fluid in low-density regions, where the absorption of a single packet is a low-probability event that radically changes the properties of the fluid in a cell. The disadvantage of that approach is that it only conserves energy, momentum and lepton number when averaging over many packet interactions. Moreover, as Monte Carlo errors scale such that if we expect $N$ neutrino-matter interactions in a region of phase space, the uncertainty on the actual number of interactions is $\sim \sqrt{N}$, absolute errors in the conservation laws are likely to grow continuously over the course of a simulation. This may be particularly problematic in high-density regions, where it is important for the simulation to properly conserve energy and momentum and interactions are extremely frequent.

To retain the advantages of the `expectation value' method in low-density regions while avoiding error buildup in high-density regions, we now use a weighted average of the two methods that smoothly transitions from the expectation value in optically thin regions to an explicit accounting of each momentum transfer event in optically thick regime. Specifically, if $G^\nu_{\rm av}$ is the momentum transfer calculated using the expectation value method (as in previous SpEC simulations) and $G^\nu_{\rm sim}$ is the momentum transfer from the actual interactions sampled in the simulation, we set
\beq
G^\nu = a G^\nu_{\rm sim} + (1-a) G^\nu_{\rm av}.
\eeq
The same weighting is applied to $C^0$. The weighting coefficient is
\beq
a = {\rm min}(1,10^{-3} \frac{M_{\rm cell}}{m_p N_p}),
\eeq
with $M_{\rm cell}$ the mass of fluid in the grid cell a neutrino packet is in, $m_p$ the proton mass, and $N_p$ the number of neutrinos in the evolved packet. This value is chosen so that the absorption of a packet representing electron (anti)neutrinos at most leads to $C^0_{\rm ex}$ changing the electron fraction of the fluid by $10^{-3}$ (this is meant to be `small enough' to limit shot noise from the use of $G^\nu_{\rm ex}$ -- but we have not so far attempted to optimize this parameter). A detailed discussion of the calculation of $G^\nu_{\rm av}$ can be found e.g. in~\cite{Foucart:2021mcb}. Calculating $G^\nu_{\rm sim}$ is actually easier. We use~\cite{Ryan2015}
\beq
\sqrt{-g} V G^\nu_{\rm sim} \Delta t = -N_p \Delta p^{\nu}
\eeq
with $\Delta p^{\nu}$ the difference between the 4-momentum of a single neutrino from the packet after an interaction (or zero for absorption) and its 4-momentum before the interaction. Here, $V$ is the coordinate volume of the grid cell the neutrino is in, $g$ is the determinant of the 4-metric, and $\Delta t$ the time step. Similarly for $C^0$, we get
\beq
\sqrt{-g} V C^0_{\rm sim} \Delta t = N_p m_p s_\nu
\eeq
for absorption, and $C^0_{\rm sim} = 0$ for scattering events. Here, $s_\nu=1$ for $\nu_e$, $-1$ for $\bar\nu_e$, and $0$ otherwise.

\subsection{Adaptive packet weights}

The number of neutrinos $N_p$ that a single packet represents is an important choice in a Monte Carlo algorithm. It is the main determinant of how we use our limited computational resources to attempt to represent the distribution function of neutrinos in a simulation. $N_p$ may vary from packet to packet. Thus, by varying $N_p$, we can focus computational resources in specific regions of phase-space. Unfortunately, there is not at this point a clear `optimal' choice for $N_p$. The best value will depend on the question that we are trying to answer. In previous simulations, we used equal-energy packet (in the fluid frame at the time of emission) -- a convenient choice to obtain rough estimates of statistical errors in the energy distribution of neutrinos. One could easily justify the use of packets representing the same number of neutrinos instead. In~\cite{Foucart:2025nub}, we argued that while these choices may be good to reduce shot noise in the coupling between neutrinos and the fluid, they are suboptimal if our objective is to accurately capture the distribution function of neutrinos $f_\nu$ in specific energy ranges. For that purpose, it may be useful for packets to represent a fixed fraction of the phase space available between given energy values, i.e. to weight packets such that the addition of a single packet in a cell modifies the estimated value of $f_\nu$ within a given grid cell and energy bin in the simulation by a constant value $\Delta f_\nu$. That choice oversamples low-energy packets compared to constant-energy or constant-number packets. Indeed, the maximum number of neutrinos of a given species in a region of phase space defined by a spatial volume $\Delta V$, energies $\in [\epsilon_0,\epsilon_1]$, and direction of propagation within a solid angle $\Delta \Omega$ is
\beq
N_{\rm max} = \frac{1}{(hc)^3} (\Delta \Omega) (\Delta V)\frac{\epsilon_1^3-\epsilon_0^3}{3}.
\eeq
Constant-energy packets have $N_p\propto \epsilon^{-1}$; constant number packets of course have constant $N_p$; packets representing a constant fraction of the available phase space (as defined above) have $N_p\propto \epsilon^3$. Calculating $f_\nu$ has not been needed in previous Monte Carlo simulations; here, however, we need $f_\nu$ for our calculations of inelastic scattering and pair production/annihilation rates -- at least in semi-transparent and optically thin regions, where we cannot reliably assume that neutrinos are in equilibrium with the fluid. As shown in~\cite{Foucart:2025nub}, constant-energy packet can practically lead to estimated values of $f_\nu \gg 1$ within low energy bins, even when a single packet is present in that cell. Here, we aim for a weighting scheme that allows us to estimate $f_\nu$ within specific energy bins, after integrating over all possible directions of propagation. As $N_p=N_{\rm max}$ corresponds to $f_\nu=1$, the addition of a single packet within a given grid cell and energy bin represents a change $\Delta f_\nu$ in our estimate of $f_\nu$ within the simulation if
\beq
N_p = g_\nu \Delta f_\nu \frac{4\pi}{(hc)^3} \Delta V \frac{\epsilon_1^3-\epsilon_0^3}{3}
\eeq
with $\epsilon_{0,1}$ now the boundary of the relevant energy bin and $g_\nu$ the number of neutrino species that the packet represents (for us, $g_\nu=1$ for electron type neutrinos and $g_\nu=4$ for heavy-lepton neutrinos, as we evolve all heavy-lepton neutrinos as a single species $\nu_x$). At the very least, we need $\Delta f_\nu <1$ to be able to calculate blocking factors $(1-f_\nu)>0$. Ideally, we would like $\Delta f_\nu \ll f_\nu$, but we will see that this is beyond what our current simulations can do.

In~\cite{Foucart:2025nub} we proposed as `optimal' for the calculation of $f_\nu$ a scheme that would effective use constant-$\Delta f_\nu$ packets. This proved impractical in our simulations without additional work. Specifically, that choice results in high-energy neutrinos being represented by very high $N_p$ packets, creating unacceptable shot noise in the coupling between the fluid and the neutrinos. Such a scheme would also require the creation of a large number of packets representing very low energy neutrinos that are generally unimportant to the evolution of the system (e.g. the lowest energy bin used in our current reaction tables is $\epsilon \in [0,4]\,{\rm MeV}$; using a large fraction of our computational resources on those neutrinos would be inefficient). Finally, that scheme would create an unsustainable number of packets in optically thick regions, where we do not need to explicitly calculate $f_\nu$ to reliably compute interaction rates: there, we can rely on equilibrium consideration instead. Fundamentally, we would like the constant-$\Delta f_\nu$ method to be used in semi-transparent and optically thin regions and for neutrinos in the most relevant $O(10\,{\rm MeV})$ energy regime, but not elsewhere. This is the main objective of our new weighting scheme.

To meet this objective, we follow the method below. We use three separate prescriptions for $N_p$:
\begin{itemize}
\item A constant energy packet prescription
\beq
N_c = \xi_s E_0/\epsilon
\eeq
with $\xi_s$ a parameter that can be varied to get the desired number of packets in a simulation (and may be different for each neutrino species), and $E_0$ a user specified reference energy.
\item A prescription meant to get a desired number of packets per grid cell in regions where neutrinos reach equilibrium with the fluid, or equivalently a fixed number of packets emitted per light crossing time of an optically thick cell. Specifically, we use
\beq
N_{\rm thick} = \frac{\eta}{N_0\epsilon}  (\sqrt{\gamma} V)^{4/3}
\eeq
with $\eta$ the total emissivity within a cell of coordinate volume $V$, and $\gamma$ the determinant of the spatial metric (the power of $4/3$ effectively accounts for the product of the cell volume and an average light crossing time of the cell). $N_0$ is a fixed user-specified variable, set to $100$ in our current simulations. With the Implicit Monte Carlo methods used in SpEC, $(\sqrt{\gamma} V)^{1/3}\sim 1/(2\kappa_a)$ in optically thick regions, and we thus end up with $\sim 2N_0$ packets per species and per cell in those regions at any given time.
\item A constant-$\Delta f$ prescription
\beq
N_f = \xi_s \Delta f_0 N_{\rm max} \xi_{corr}
\eeq
with $\xi_s$ the same scaling parameter as for $N_c$ and $\Delta f_0$ a user specified reference `fraction of $f_\nu$'. To avoid issues with overresolution of the lowest energy bin, we use the prescription from the second-lowest energy bin for the two lowest energy bins. The parameter $\xi_{\rm corr}\geq 1$ is set in such a way that $N_f$ is large in dense regions (the scheme will then default to one of the other two estimates). Our current prescription is
\beq
\xi_{\rm corr} = {\rm max}\left(1,\frac{\rho}{10^{13}\,{\rm g/cc}}\right).
\eeq
In addition, to force the use of $N_{\rm thick}$ in optically thick regions, we use $N_f=N_{\rm thick}$ when the optical depth of a cell (defined using the thermalization opacity $\kappa_{\rm th}=\sqrt{\kappa_a (\kappa_a+\kappa_s)}$) is larger than $10$, with a linear transition between $N_{\rm thick}$ and $N_f$ for optical depths between $5$ and $10$. We note that we use here the optical depth across a given cell, not the total optical depth between a grid cell and the domain boundary.
\end{itemize}
We then combine these three prescriptions using
\beq
N_p = {\rm min}(N_{f},{\rm max}[N_{\rm thick},N_c]).
\eeq
The parameter $\xi_s$ is adjusted on the fly, as in earlier simulations, to reach a desired number of Monte Carlo packets. We also reevaluate the desired number of neutrinos in each packet after each step. If the desired $N_p$ is more than twice the actual $N_p$, we down-sample the packet: it has a $50\%$ chance of being destroyed, but if it survives it will represent twice as many neutrinos. If the desired $N_p$ is less than half the actual $N_p$, we split the packet in two. This obviously does not create two independent samples, but this operation generally takes place in dense regions where the packets will quickly be absorbed or scattered, which mitigates that issue. Earlier simulations effectively used $N_p = {\rm max}[N_{\rm thick},N_c]$ with $N_0=100$ and $E_0=10^{-12}M_\odot c^2$, and this prescription is also used in this work for simulations that do not use our new adaptive weighting of Monte Carlo packets.

If we consider the different regions of a simulation:
\begin{itemize}
\item In optically thin regions, we will have $N_{\rm thick}<N_c$. Packets representing low-energy neutrinos will use $N_f$ (constant $\Delta f_\nu$), while packets representing high-energy neutrinos will use $N_c$ (constant energy). This avoids excessive shot noise from high-energy neutrino packets while allowing us to better resolve $f_\nu$ for low energy neutrinos.
\item In optically thick regions, we will use $N_{\rm thick}$
\item When a packet crosses a refinement boundary, it will be down-sampled, thus avoiding the propagation of a large number of packets in outer regions where neutrino-matter interactions are uncommon.
\end{itemize}

This method has a number of free parameters that one could tune to optimize performance. In particular, the ratio of $E_0$ and $\Delta f_0$ determines where we use the `constant $\Delta f$' prescription in optically thin regions, while $N_0$ will set where we switch to a constant number of packets per cell. In the simulations presented here, we used $\Delta f_0=10^{-3}$ and $E_0=10^{-12}M_\odot c^2$. These values were chosen because earlier simulations in which we attempted to use only $N_f$ indicated that it would lead to marginally resolving $f_\nu$ -- which we know to be the best case scenario for our current number of packets~\cite{Foucart:2025nub}, while simulations that did not use $N_f$ ended with packets of energy $\sim 3\,10^{-10}M_\odot c^2$ just before collapse of the remnant to a black hole (in simulation M136-M126, described below). We will see that in our simulations using the new weights, we get $\Delta f\sim 0.01-0.6$ depending on the species and the exact time after merger, and thus packet energies of twice what we used in the old scheme; this choice thus attempts to distribute our resource equally between `packets limited by $\Delta f_\nu$' and `packets limited by their total energy'.

There are nearly certainly ways to improve upon this. In particular, one might wonder whether $N_0$ could not be drastically smaller in optically thick regions far from the neutrinosphere, where neutrinos are in equilibrium with the fluid anyways. We only tried to do this in simulations performed before we changed the method used to calculate the coupling between neutrinos and the fluid, and this resulted in unacceptable shot noise and eventual instabilities in the denser regions of the star. At this point, it is unclear whether our new algorithm for neutrino-matter coupling would allow us to use a lower $N_0$. It would certainly be desirable, considering that regions where the fluid and neutrinos are in equilibrium currently host a majority of the neutrino packets in our simulations, just to model neutrinos near statistical equilibrium with the fluid (although~\cite{Alford:2026kwd} showed that even at temperatures $\sim 20\,{\rm MeV}$ deviations from equilibrium remain important in the calculation of interaction rates). Another potential optimization would be to use the same $\xi_s$ for all species, instead of fixing the number of packets per species. We have not so far attempted to do this.

\subsection{Inelastic scattering on $e^-$}

The scattering of neutrinos and antineutrinos on $e^-$ typically involves more significant energy transfers between neutrinos and $e^-$ than scattering on nucleons and nuclei. The contribution of these events to the evolution of the distribution function $f$ is~\cite{1985ApJS...58..771B}
\beqn
&&\frac{\partial f}{\partial t}(\epsilon,\Omega)\\
 &=& \frac{1-f(\epsilon,\Omega)}{(h c)^3}
\int d\epsilon' (\epsilon')^2 \int d\Omega' f(\epsilon',\Omega') R^{in}(\epsilon,\epsilon',\mu)\nonumber\\
&&- \frac{f(\epsilon,\Omega)}{(h c)^3}\int d\epsilon' (\epsilon')^2 \int d\Omega' [1-f(\epsilon',\Omega')] R^{out}(\epsilon,\epsilon',\mu)\nonumber
\eeqn
with $\epsilon$ the energy of a neutrino, $\Omega$ its direction of propagation on the unit sphere, and $\mu=\cos\theta$ with $\theta$ the angle between the direction of propagation of the neutrinos. Primed quantities refers to the incoming (outgoing) neutrino for the emission (absorption) term. In the NuLib library used in our simulations~\cite{OConnor:2015}, the kernels $R^{\rm in/out}$ are expended in Legendre polynomials in $\mu$ and tabulated in $\epsilon,\epsilon'$. For this first attempt to include inelastic scattering in our simulations, we note that the rate of scattering onto electrons is, in regions where these events most impact the evolution of neutron star mergers, subdominant with respect to near-elastic scattering on nuclei~\cite{Rath:2026vfr}. As a result, we expect that in any given region of the simulation, either inelastic scattering on electrons is negligible or the distribution function of neutrinos is close to isotropic. We thus only keep the zeroth-order term of the expension of the kernels in $\mu$ i.e., in the notation of~\cite{1985ApJS...58..771B} and NuLib,
\beq
R^{\rm in/out}(\epsilon,\epsilon',\mu) = \frac{1}{2} \Phi_0^{\rm in/out}(\epsilon,\epsilon').
\eeq
This of course greatly simplifies the integrals,
\beqn
&&\frac{\partial f}{\partial t}(\epsilon,\Omega)\\
 &=& \frac{1-f(\epsilon,\Omega)}{(h c)^3}
\int d\epsilon' (\epsilon')^2  \langle f(\epsilon')\rangle (2\pi) \Phi_0^{in}(\epsilon,\epsilon')\nonumber\\
&&- \frac{f(\epsilon,\Omega)}{(h c)^3}\int d\epsilon' (\epsilon')^2  [1-\langle f(\epsilon')\rangle] (2\pi) \Phi_0^{out}(\epsilon,\epsilon')\nonumber
\eeqn
with $\langle f(\epsilon) \rangle$ the average value of $f$ for neutrinos of energy $\epsilon$. Practically, we evaluate the first integral as
\beqn
\int d\epsilon' (\epsilon')^2  \langle f(\epsilon')\rangle \Phi_0^{in}(\epsilon,\epsilon') =\\
\sum_{i=0}^{N-1} \tilde f_{i}  \frac{\epsilon^3_{i+1/2}-\epsilon^3_{i-1/2}}{3}  \Phi_0^{in}(\epsilon,\epsilon_{i}')
\eeqn
with the sum taken over all energy groups in our NuLib table. Here $\epsilon_{i\pm 1/2}$ are the bounds of the energy groups, $\epsilon_{i}$ the central energy of each group, and $\tilde f_{i}$ is the average value of $f$ within an energy group. Similarly,
\beqn
\int d\epsilon' (\epsilon')^2  [1-\langle f(\epsilon')\rangle] \Phi_0^{out}(\epsilon,\epsilon') =\\
\sum_{i=0}^{N-1}(1- \tilde f_{i})  \frac{\epsilon^3_{i+1/2}-\epsilon^3_{i-1/2}}{3}  \Phi_0^{out}(\epsilon,\epsilon_{i}').
\eeqn
We can then write
\beq
\partial_t f(\epsilon,\mu,\phi) = \Gamma^+ - \Gamma^- f(\epsilon,\Omega)
\eeq
with the computation of $\Gamma^\pm$ only requiring an estimate of $\tilde f_i$ as well as $\Phi_0^{in/out}(\epsilon,\epsilon_{i}')$. To evaluate $\Phi_0$, we use linear interpolation from the NuLib table in log-density, log-temperature, electron fraction, and log-$\epsilon$. 
For $\tilde f_i$, we use
\beq
\tilde f_{i} = \frac{N_{\rm est,i}}{N_{\rm max,i}}
\eeq
with $N_{\rm est,i}$ the estimated number of neutrinos within the current grid cell and energy bin in the Monte Carlo simulation, and $N_{\rm max}$ the number of neutrinos that would be present in that energy bin and grid cell if we had $f=1$. In optically thick regions, we set $N_{\rm est}$ to its {\it expected equilibrium value} $N_{\rm eq}$, as using those values is more accurate than an estimate subject to Monte Carlo sampling noise. In optically thin regions, we use the average number of neutrinos within the Monte Carlo simulation during the last time step, $N_{\rm sim}$ (calculated from the average energy density of neutrinos within a given energy bin and grid cell, estimated as described in~\cite{Foucart:2021mcb}). Specifically, we use
\beq
N_{\rm est,i} = \alpha N_{\rm eq} + (1-\alpha) N_{\rm sim}
\eeq
where $\alpha=1$ if the lengthscale $1/\sqrt{\kappa_a(\kappa_a + \kappa_s)}$ (ignoring inelastic scattering) is shorter than $100\,{\rm m}$, $\alpha=0$ if it is larger than $125\,{\rm m}$, and we interpolate linearly between $0$ and $1$ in between those two lengthscales. Thus we only ignore the number of neutrinos obtained from the simulation if the thermalization timescale of neutrinos is shorter than the light-crossing time of our grid cells.

The fact that the only information about $f$ needed to calculate interaction rates is $\tilde f_{i}$ is an important motivation for our choice of method. The emissivity $\eta_i$ in bin $i$ and absorption opacity $\kappa_a$ typically used in SpEC are then
\beq
\kappa_a (\epsilon) = \Gamma^- (\epsilon)/c;\,\,\eta_i = \pi (\epsilon_{i+1/2}^4-\epsilon_{i-1/2}^4)\frac{\Gamma^+}{(hc)^3}
\eeq
with $\eta_i$ the isotropic emissivity integrated over an energy bin and all emission angles. We note that, in this formalism, there is no blocking factor involved in the calculation of $\eta_i$; all terms proportional to $f$ have been moved to the absorption, as when using the `stimulated absorption' method for charged current reaction.

One limitation of this formalism is worth noting:  by writing the contribution of inelastic scattering as an `absorption' and `emission' term rather than individual neutrino packets being scattered inelastically, we can no longer guarantee that inelastic scattering has a net zero effect on the evolution of the total electron lepton number to roundoff accuracy. Instead, it does so only within the `discretization error' of our Monte Carlo simulation (sampling noise, split time step,...). Considering that inelastic scattering is a subdominant effect for $\nu_e,\bar\nu_e$ that mainly contributes to thermalizing $\bar\nu_e$ close to the neutrinosphere without modifying the equilibrium distribution of neutrinos in dense regions, this is likely a minor source of error in the simulation -- but certainly something that could be improved upon.

\subsection{Pair creation and annihilation}

Our treatment of the reaction $\nu + \bar \nu \leftrightarrow e^+ + e^-$ is similar to our treatment of inelastic scattering, though with more attention paid to the relative orientation of the neutrinos and antineutrinos in the forward reaction. Indeed, neutrino-antineutrino pair annihilation is potentially an important source of energy deposition in low-density polar regions, where the neutrino distribution is definitely not isotropic; and the reaction cross-section is sensitive to the angle between the momenta of the (anti)neutrinos.

We have for pair creation/annihilation,
\beqn
&&\frac{\partial f}{\partial t}(\epsilon,\Omega) =\\
&& \frac{1-f(\epsilon,\Omega)}{(h c)^3}
\int d\epsilon' (\epsilon')^2 \int d\Omega' [1-\bar f(\epsilon',\Omega')] R^{p}(\epsilon,\epsilon',\mu)\nonumber\\
&&- \frac{f(\epsilon,\Omega)}{(h c)^3}\int d\epsilon' (\epsilon')^2 \int d\Omega' \bar f(\epsilon',\Omega') R^{a}(\epsilon,\epsilon',\mu)\nonumber
\eeqn
with the same notation as in the previous section and using $\bar f$ for the distribution function of antiparticles. In order to estimate these integrals without requiring more fine grained information about $\bar f$ than for inelastic scattering, we approximate them as
\beqn
&&\int d\epsilon' (\epsilon')^2 \int d\Omega' [1-\bar f(\epsilon',\Omega')] R^{p}(\epsilon,\epsilon',\mu) =\nonumber\\
&&\sum_i (1-\bar f_i) \Phi_0^p(\epsilon,\epsilon') (2\pi)\frac{\epsilon^3_{i+1/2}-\epsilon^3_{i-1/2}}{3}
\eeqn
and
\beqn
&&\int d\epsilon' (\epsilon')^2 \int d\Omega' \bar f(\epsilon',\Omega') R^{a}(\epsilon,\epsilon',\mu) =\nonumber\\
&&\sum_i \bar f_i \Phi_0^a(\epsilon,\epsilon') (2\pi)\frac{\epsilon^3_{i+1/2}-\epsilon^3_{i-1/2}}{3} \alpha_{\rm form}.
\eeqn
Here $\alpha_{\rm form}$ is a geometrical form factor capturing the correction to the annihilation cross section due to the relative orientation of the interacting particles {\it averaged over all packets within a cell during the previous timestep, and ignoring blocking factors}. The motivation for this approximation is that for the first integral ($\nu\bar\nu$ pair creation), an isotropic distribution is a reasonable approximation for $\bar f_i$ in regions whenever $(1-\bar f_i)$ is not negligible. For the second, we assume effectively that the distribution functions are separable, $f(\epsilon,\mu,\phi) \approx f_1(\epsilon) f_2(\mu,\phi)$. This is definitely a source of error: high energy and low energy neutrinos have very different decoupling surfaces and thus will have different angular distributions in optically thin regions. It does however allow us to capture the average effect of the angular distribution of neutrinos. Based on the expression of the pair annihilation rate in low-density regions expressed as a function of the moments of the neutrino distribution function described in~\cite{Fujibayashi:2017abc}, we use the form factor
\beq
\alpha_{\rm form} = \frac{T_{\mu\nu} \bar T^{\mu\nu}}{T_{\alpha\beta}^{\rm iso} \bar T^{\alpha\beta}_{\rm iso}}.
\eeq
Here the stress-energy tensor of neutrinos $T_{\mu\nu}$ is calculated from the average moments of the distribution function of neutrinos over the last time step, while $T_{\alpha\beta}^{\rm iso}$ is its value for the same fluid-frame energy density but assuming an isotropic distribution function in the fluid frame. This formula is correct in low density regions and for a single energy~\cite{Fujibayashi:2017abc}. Proceeding exactly as for inelastic scattering, we can then calculate coefficients $\Gamma^\pm$ such that
\beq
\partial_t f = \Gamma^+ - \Gamma^- f
\eeq
and relate them to an effective emissivity and absorption opacity for neutrinos. We compute $\Gamma^\pm$ in two ways: first using the values of $\bar f_i$ and $\alpha_{\rm form}$ measured in our simulation, then using $\bar f_i=\bar f_{\rm eq}$ and $\alpha_{\rm form}=1$. We then linearly interpolate between those two values as when estimating $N_{\rm est,i}$ in the previous section (i.e. using the equilibrium values when the thermalization lengthscale is shorter than $100\,{\rm m}$).

As for inelastic scattering, there is a clear drawback to treating pair processes as emission/absorption coefficients in this way: there is no guarantee that we create/annihilate exactly as many neutrinos as antineutrinos. It does, however, allow us to approximately capture pair processes in both optically thick and optically thin regions, partially accounting for both blocking factors and the geometry of the neutrino distribution function.

\section{Simulation setup}
\label{sec:sim}

\subsection{Initial Data}

\begin{table}
\begin{tabular}{c|ccccc}
Name & $M_1$ & $M_2$ & Weight & Kernels & $t_{\rm BH}$\\
\hline
M140-M130 & $1.40M_\odot$ & $1.30M_\odot$ & Fixed & No & $4\,{\rm ms}$\\
M136-M126 & $1.36M_\odot$ & $1.26M_\odot$ & Fixed & No & $6\,{\rm ms}$\\
M127-M118-Base & $1.27M_\odot$ & $1.18M_\odot$ & Fixed & No & --\\
M127-M118-Adapt & $1.27M_\odot$ & $1.18M_\odot$ & Adaptive & No & --\\
M127-M118-Ker & $1.27M_\odot$ & $1.18M_\odot$ & Adaptive & Yes & --\\
\end{tabular}
\caption{Summary of the configurations simulated in this manuscript. We list the name of the simulation, the gravitational mass of each neutron star, weight used for the Monte Carlo packets (`Fixed' when all packets have the same energy at a given time step, `Adaptive' for the scheme described in the text that aims to resolve the distribution function of neutrinos), and time between merger and collapse to a black hole. When no time in provided, a neutron star survives at least $10\,{\rm ms}$.}
\label{tab:id}
\end{table}

The initial data for the simulations performed here is generated with the Spells code~\cite{Pfeiffer2003,FoucartEtAl:2008}. All simulations use the SFHo equation of state~\cite{Steiner:2012rk}, with initial temperature $T=0.1\,{\rm MeV}$ and initial electron fraction $Y_e$ set by requiring neutrinoless beta-equilibrium. We perform one round of eccentricity reduction for each system. We consider three different initial conditions. The first is the system already published in~\cite{Foucart:2024npn}: a binary neutron star system with component gravitational massees $M_1=1.4M_\odot$ and $M_2=1.3M_\odot$ and initial separation of $45\,{\rm km}$. To study systems with a slightly longer lived neutron star, we then reduced the mass of both neutron stars by first $3\%$, then $10\%$. The initial separation is reduced by the same fraction, to maintain roughly the same number of orbits to merger. The latter configuration is the main focus of this manuscript, and was simulated with three distinct setups for the treatment of neutrinos. We use the mass of the neutron stars as the basis for the naming conventions of those three configurations: M140-M130, M136-M126, M127-M118.

We note that the maximum mass of an isolated, non-rotating neutron star is $M_{\rm max}=2.06M_\odot$ for SFHo, while the threshold mass for prompt collapse was previously estimated as $M_{\rm thresh}=2.95M_\odot$~\cite{PhysRevLett.111.131101}. Our systems thus have total mass $M_{\rm tot}\sim (1.2-1.3)M_{\rm max}$, where we would expect neutron star remnant that require differential rotation and/or thermal support to avoid collapse.

\subsection{Evolution}

Each of our three configurations is evolved using the SpEC code and the same numerical methods as in~\cite{Foucart:2024npn}. Briefly, the metric is evolved on a pseudospectral grid using the generalized harmonic formulation of Einstein's equation~\cite{Lindblom:2007} in the harmonic gauge and the same grid structure as in~\cite{Foucart:2024npn}. The neutron star is modeled as an ideal fluid, with the relativistic fluid equations evolved using finite volume methods and high-order shock capturing methods described in~\cite{Duez:2008rb,Foucart:2013a}. We use fixed mesh refinement with a finest grid at resolution $\Delta x \approx 180\,{\rm m}$ during merger. That grid has $256^3$ cells. Coarser grids have grid spacing increased in factors of $2$, and we use four mesh refinement levels in total. The coarsest grid thus expands $\sim 170\,{\rm km}$ from the center of mass of the system. 
We evolve all configurations up to at least $10\,{\rm ms}$ post-merger.

Neutrinos are evolved using our energy-dependent Monte Carlo transport code~\cite{Foucart:2021mcb}, with a target of $10^8$ packets per species. We consider only $3$ distinct species: $\nu_e$, $\bar \nu_e$, and a heavy-lepton neutrinos species $\nu_x$ that includes the four other species of (anti)neutrinos. All three configurations are first evolved using our `base' methods (i.e. without reaction kernels) and the same set of interactions as in our previous Monte Carlo simulations: charged current reactions $n+e^+ \leftrightarrow p + \bar\nu_e$ and $p+e^- \leftrightarrow n + \nu_e$; scattering on neutrons, protons, alpha particles and heavy nuclei; and for heavy-lepton neutrinos only, $\nu\bar\nu \leftrightarrow e^+ e^-$ and nucleon-nucleon Bremsstrahlung. We use the NuLib library~\cite{OConnor:2015} to generate an emission rate $\eta$ as well as absorption and scattering opacities $\kappa_{a,s}$. We note that scattering is assumed to be elastic and isotropic. More importantly, for pair processes (the last two reactions), we use emission rates assuming an equilibrium distribution of neutrinos for the blocking factors, then an absorption rate using Kirchoff's law $\eta = f_{\nu,eq} \kappa_a$. This is a very significant approximation for those rates, mainly aimed at guaranteeing that heavy lepton neutrinos are in equilibrium with the fluid in dense regions.

The last configuration, M127-M118, is simulated with three different treatments of neutrinos. We refer to the simulation using the reactions described in the previous paragraph as M127-M118-Base. We then perform a simulation with the adaptive weighting of the neutrino packets and improvements to the treatment of neutrino-matter couplings, but without changing the interaction rates (M127-M118-Adapt). Finally, we use our latest weights in combination with our improved treatment of $\nu\bar\nu \leftrightarrow e^+ e^-$ and with the inclusion of inelastic scattering on electrons, within the approximations described in previous sections (called M127-M118-Ker). For that last simulation, we use NuLib interaction kernels instead of emission/absorption rates. We note that we also performed a number of simulations with different weighting schemes that attempted to capture the distribution function of neutrinos but ended up with weights on high-energy packets (or packets in very dense regions) that were high enough to create significant numerical errors. We do not report on these failed simulations in detail here. Most of this manuscript focuses on a comparison between M127-M118-Adapt and M127-M118-Ker.

It is worth noting that some potentially important neutrino physics is ignored in all of these simualations. This includes for example charged-current reactions involving muons, neutrino flavor conversion, and energy exchanges in the scattering of neutrinos on nucleons. The first two have been studied using more approximate transport schemes in the context of neutron star mergers~\cite{Li:2021vqj,Ng:2024zve,Gieg:2026beb,Qiu:2025kgy}, while the latter can have an effect on the thermalization of neutrinos comparable to scattering on electrons in core-collapse supernovae~\cite{Janka:1995ir,PhysRevC.62.035802,Burrows:2016ohd} but has not been studied in the context of neutron star mergers. The impact of inelastic scattering of neutrinos on electrons observed in our simulations thus likely underestimates the total impact of inelastic scattering of neutrinos in neutron star mergers.

\section{Results}
\label{sec:results}

\subsection{Impact of the total mass of the binary on black hole collapse}

\begin{figure}
\includegraphics[width=0.9\columnwidth]{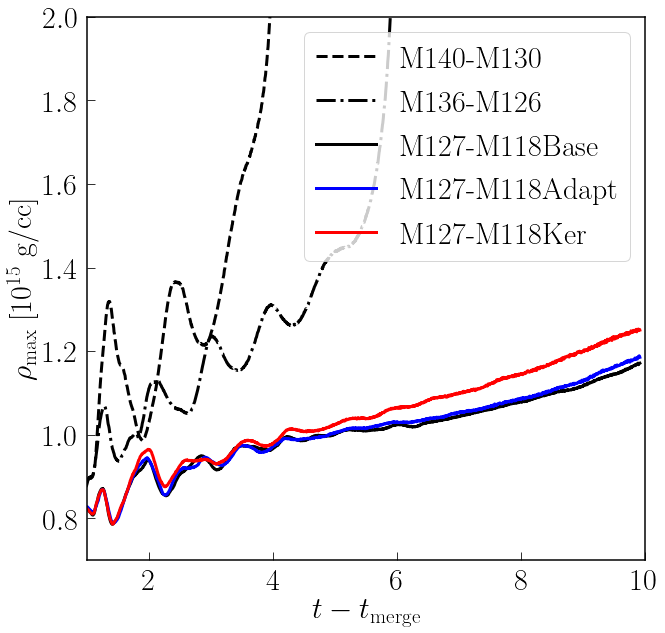}
 \caption{Maximum baryon density on the computational domain as a function of the time post-merger for the 5 simulations of Table~\ref{tab:id}, up to collapse to a black hole.}
\label{fig:maxrho}
\end{figure}

\begin{figure}
\includegraphics[width=0.9\columnwidth]{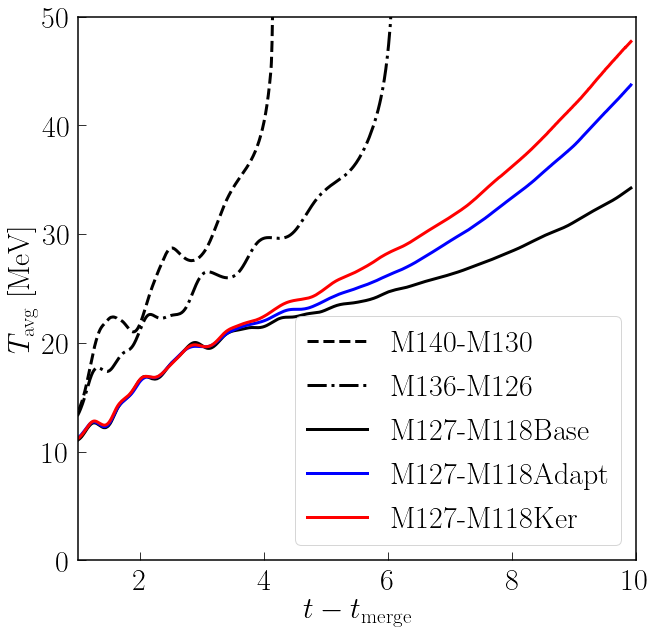}
 \caption{Mass-weighted average temperature on the computational domain as a function of the time post-merger for the 5 simulations of Table~\ref{tab:id}, up to collapse to a black hole.}
\label{fig:avgT}
\end{figure}

We first begin with a broad overview of the evolution of the remnant. One of our objectives in varying the mass of the initial neutron stars was to study the impact of that choice on the collapse time of the remnant, as well as to obtain configurations that collapse over $O(10\,{\rm ms})$ -- as such remnants will eventually allow us to study the impact of neutrinos on neutron star-disk systems and the long term evolution of black hole-disk systems. We note that there is no unique definition of the `time of merger' in these systems. Here, we use the time at which the maximum density on the grid first rises by more than $3\%$ above its value at the beginning of the simulation (i.e. when the two cores collide). The variation of the collapse time with the total mass of the binary is most easily seen on Fig.~\ref{fig:maxrho}, which shows the maximum baryon density on the grid for all five simulations. We see that the collapse time does not change much between M140-M130 ($4\,{\rm ms}$) and M136-M126 ($6\,{\rm ms}$). For reference, this is less than the change observed when varying the neutrino transport scheme used in the M140-M130 simulation, as discussed in~\cite{Foucart:2024npn} (M140-M130 evolved with a gray two-moment scheme collapsed after $\sim 8\,{\rm ms}$). M127-M118 does not collapse over the $10\,{\rm ms}$ of post-merger evolution performed here, but the steady increase in the maximum density indicates that it is likely close to black hole formation by the end of the evolution. The treatment of neutrinos within the simulation only has a minor impact on this observable. We note that this was not the case in simulations that used Monte Carlo packets with significantly higher weights in high-density regions and/or for high-energy neutrinos. Then, numerical errors led to a more rapid increase of the maximum density, indicating that large Monte Carlo errors can impact the evolution of the high-density regions of post-merger remnants.
The details of neutrino transport has a more visible effect on the mass-weighted average temperature of the remnant, shown on Fig.~\ref{fig:avgT}. There, we can see that the two simulations using adaptive packet weights (and thus practically a lower number of packets in dense regions) lead to additional heating or less efficient cooling in the bulk of the neutron star remnant as it approaches collapse.

\subsection{Matter outflows}

\begin{table}
\begin{tabular}{c|ccc}
Name & $M_{\rm unbound}$ & $\langle Y_e \rangle$ &  $\langle s \rangle$\\
\hline
M140-M130 & $0.0083M_\odot$ & 0.224 & 15.4\\
M136-M126 & $0.0060M_\odot$ &  0.233 & 16.5\\
M127-M118-Adapt & $0.0027M_\odot$ & 0.203 & 17.0\\
M127-M118-Ker & $0.0041M_\odot$ & 0.204 & 16.2\\
\end{tabular}
\caption{Summary of the properties of the unbound matter which has left the computational grid $10\,{\rm ms}$ post-merger. We show the total mass of unbound outflows, as well as their average electron fraction and specific entropy (in units of $k_b$ per baryon).}
\label{tab:outflows}
\end{table}

We now turn to the matter ejected during and immediately after merger. A summary of the global properties of the outflows at $10\,{\rm ms}$ post-merger is presented in Table~\ref{tab:outflows}. For this section, we note that M127-M118Base was evolved with a larger domain,  by which point the equation of state table was truncated at a density too high to reliably measure outflows beyond about $5\,{\rm ms}$ post-merger, and is thus not used in our study of unbound matter. 

The global properties of the outflows are listed in Table~\ref{tab:outflows}. For all simulations, we consider the unbound criterium from~\cite{Foucart:2021ikp} that accounts for heating from r-process nucleosynthesis and cooling from neutrino emission. This criteria predicts an asymptotic Lorentz factor $W_{\infty}$, from which an asymptotic velocity $v_\infty$ can be recovered. We look at all unbound matter leaving the outer boundary of our simulations in the first $10\,{\rm ms}$ of post-merger evolution. We find that the more massive systems eject significantly more material during that time -- from $0.0083M_\odot$ for M140-M130 down to $0.0027M_\odot$ for M127-M118-Adapt. This is not overly surprising. We will see below that most of the ejected material in the higher mass systems is hot, high-$Y_e$ ejecta typically associated with core-bounce oscillations of the post-merger remnant. Fig~\ref{fig:maxrho} clearly shows that these oscillations are stronger for the high mass, rapidly collapsing systems. It is worth noting however that by $10\,{\rm ms}$ post-merger, the higher mass system has largely completed its mass ejection. Less than $0.001M_\odot$ of material remains on the grid, and about $0.002M_\odot$ of bound material has left the grid, and would slowly fall back on the low-mass accretion disk formed around the remnant black hole. For M136-M126, we continued to evolution to $14\,{\rm ms}$ post-merger, by which point it reaches a similar state. By that time, a total of $0.0067M_\odot$ of unbound matter and $0.0033M_\odot$ of bound matter has left the grid. For M127-M118, on the other hand, mass ejection is likely to continue up to a few milliseconds past black hole collapse. At $10\,{\rm ms}$ post-merger, about $0.001M_\odot/{\rm ms}$ of unbound matter is still leaving the computational grid. M127-M118-Adapt would thus likely end up much closer to the total mass ejection of M136-M126 if those late-time outflows were accounted for. For all simulations, the average entropy is $\sim (15-16)k_B$ per baryon, and the average electron fraction $0.20-0.23$.

In terms of the impact of neutrino physics, we note that M127-M118-Ker ejects $50\%$ more unbound material than M127-M118-Adapt during the first $10\,{\rm ms}$ of post-merger evolution. This is a fairly significant difference, likely due to enhanced energy deposition from heavy-lepton neutrinos in regions where outflows are generated.

\begin{figure}
\includegraphics[width=0.9\columnwidth]{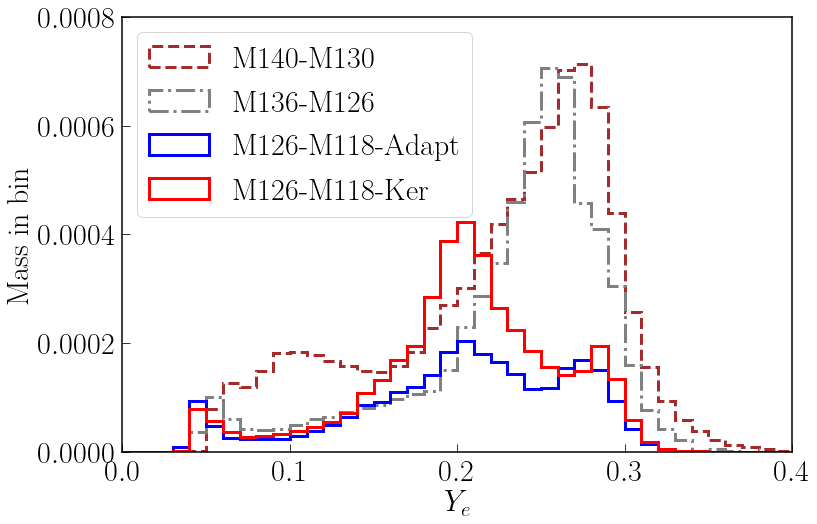}
 \caption{Electron fraction distribution of the unbound mass for each simulation.}
\label{fig:histye}
\end{figure}

The electron fraction of the ejecta also varies between systems. In Fig.~\ref{fig:histye}, we show histograms of the electron fraction for each simulation. We see that the highest mass system appears to have two main components: high-$Y_e$ outflows ($Y_e>0.25$) likely associated with core-bounce, and low-$Y_e$ outflows ($Y_e\sim 0.1$) likely associated with tidal disruption. The intermediate mass system has much less tidal ejecta. The two low-mass systems largely eject mass around a third peak at $Y_e\sim 0.2$, which is likely associated with the longer evolution of the neutron star-disk system, e.g. the spiral-arm driven ejecta described in~\cite{Radice:2023xxn}. This is the component most impacted by neutrino physics: we see that M127-M118Adapt and M127-M118Ker have similar amounts of mass in the low-$Y_e$ and high-$Y_e$ peaks, but M127-M118Ker produces much more outflows at $Y_e\sim 0.2$.

\begin{figure*}
\includegraphics[width=0.345\textwidth]{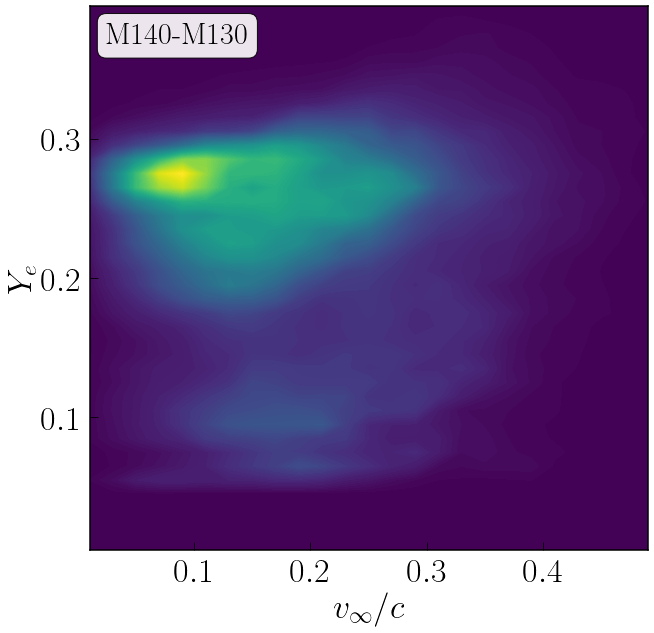}
\includegraphics[width=0.302\textwidth]{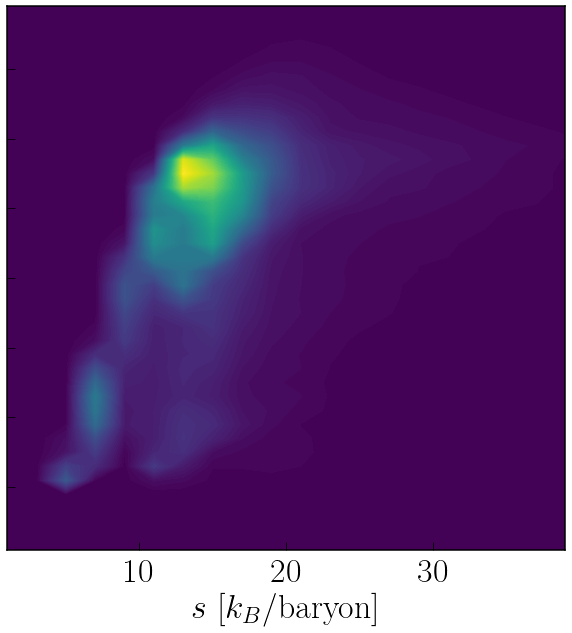}
\includegraphics[width=0.31\textwidth]{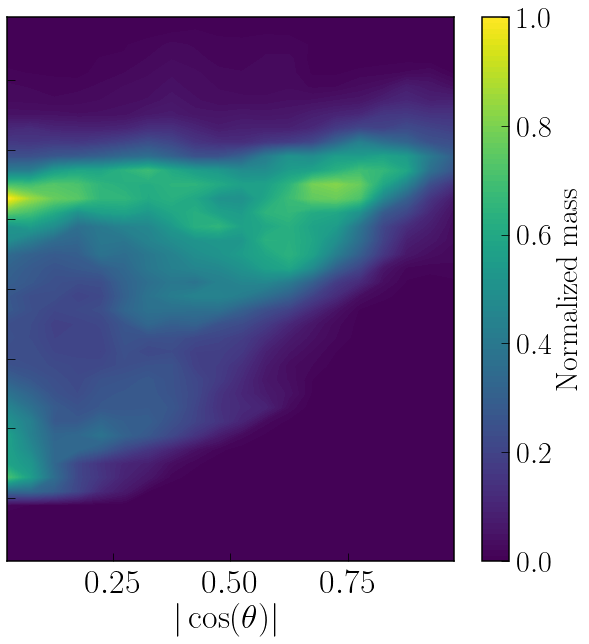}
 \caption{Distribution of the unbound mass $10\,{\rm ms}$ post-merger for simulation M140-M130. {\it Right}: Electron fraction against asymptotic velocity; {\it Middle}: Electron fraction against specific entropy; {\it Right}: Electron fraction against azimuthal angle. We use bins of $\Delta Y_e=0.01$, $\Delta v_{\infty}=0.02c$, $\Delta s=2$, and $\Delta \cos{\theta}=0.05$.}
\label{fig:M02D}
\end{figure*}

\begin{figure*}
\includegraphics[width=0.345\textwidth]{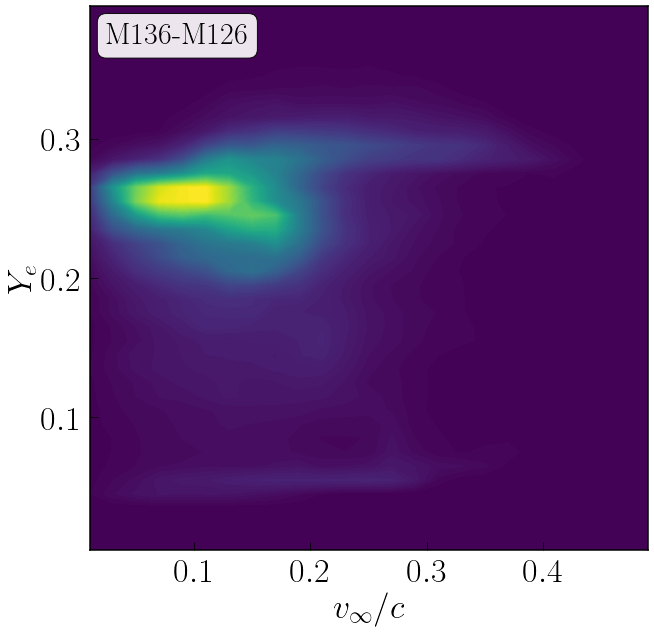}
\includegraphics[width=0.302\textwidth]{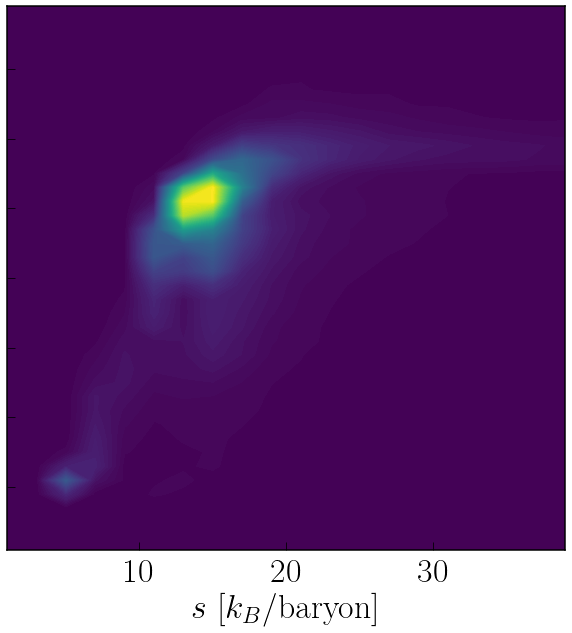}
\includegraphics[width=0.31\textwidth]{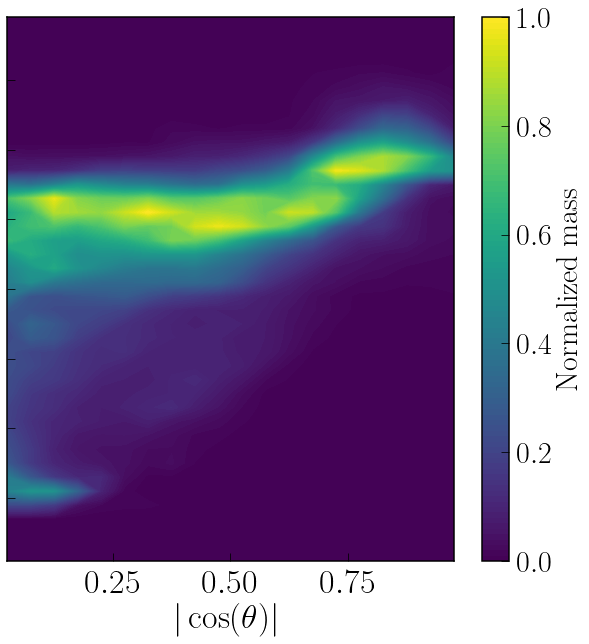}
 \caption{Same as Fig.~\ref{fig:M02D}, but for simulation M136-M126.}
\label{fig:M32D}
\end{figure*}

\begin{figure*}
\includegraphics[width=0.345\textwidth]{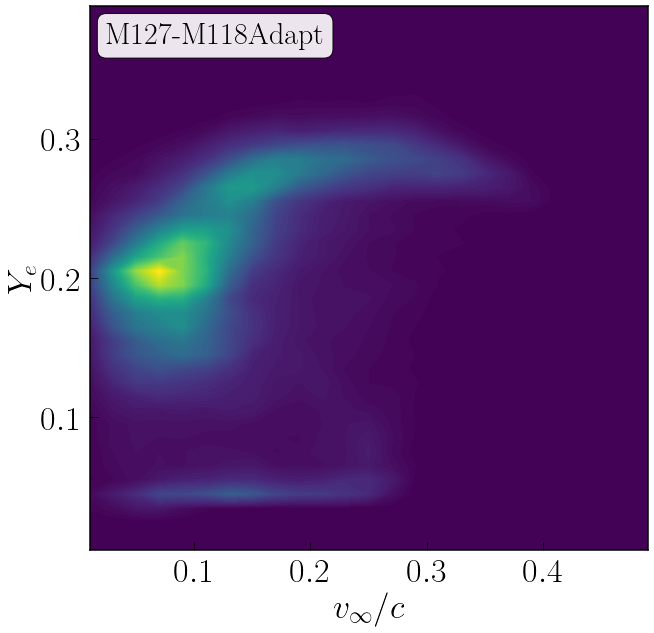}
\includegraphics[width=0.302\textwidth]{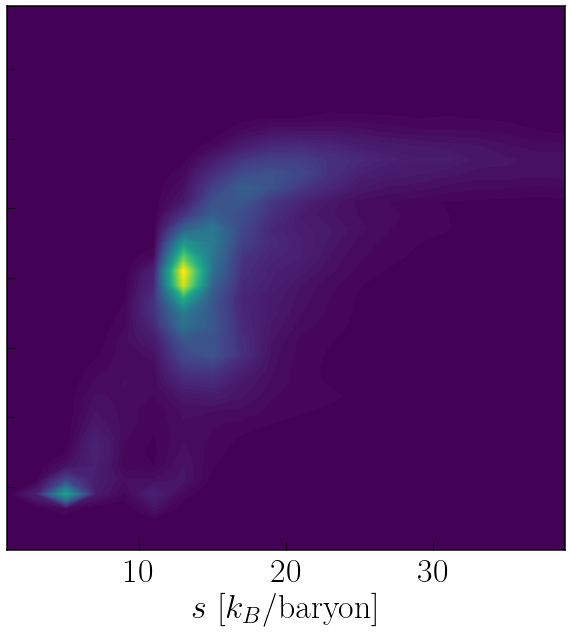}
\includegraphics[width=0.31\textwidth]{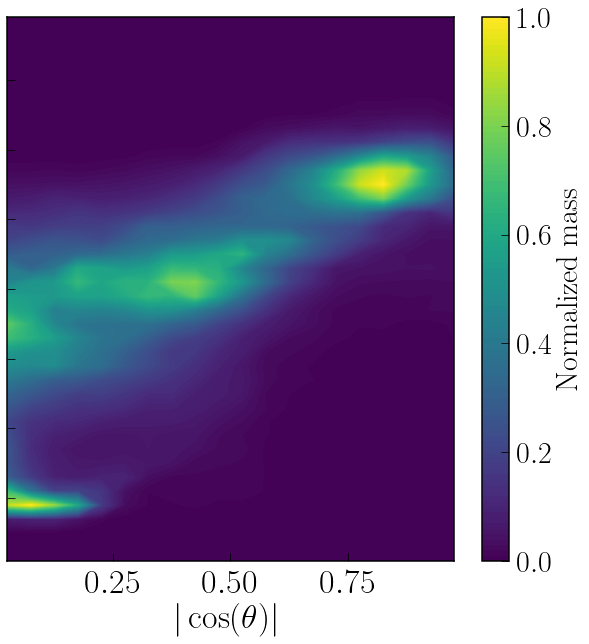}
 \caption{Same as Fig.~\ref{fig:M02D}, but for simulation M127-M118Adapt.}
\label{fig:M10a2D}
\end{figure*}

\begin{figure*}
\includegraphics[width=0.345\textwidth]{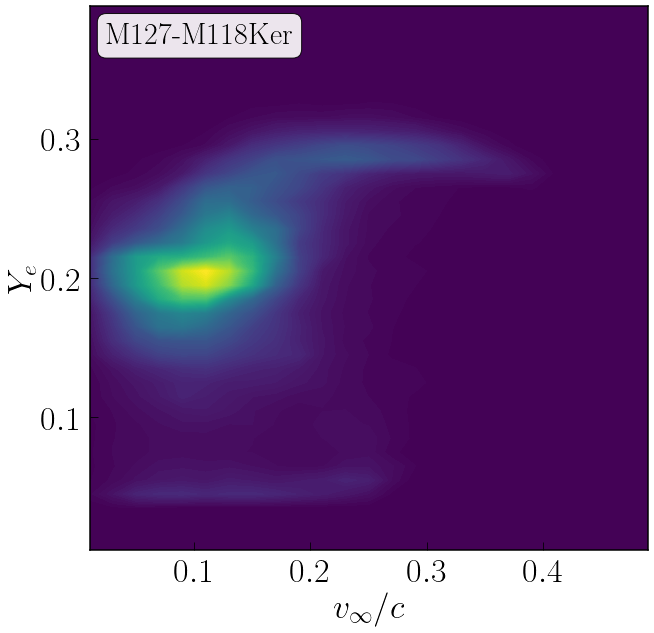}
\includegraphics[width=0.302\textwidth]{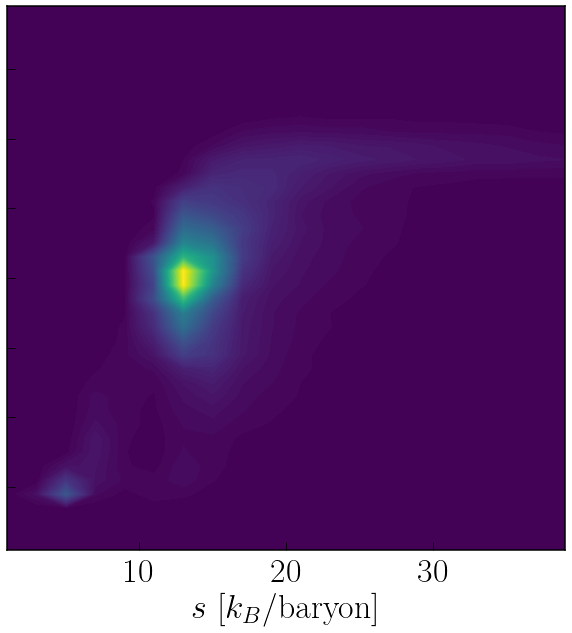}
\includegraphics[width=0.31\textwidth]{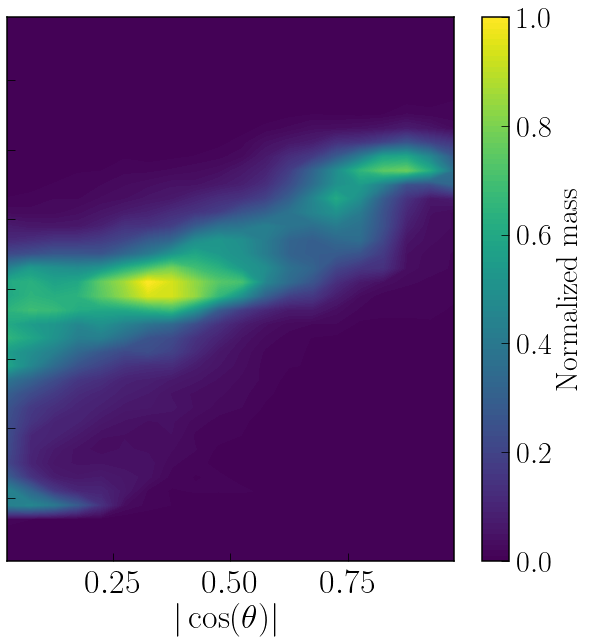}
 \caption{Same as Fig.~\ref{fig:M02D}, but for simulation M127-M118Ker.}
\label{fig:M10k2D}
\end{figure*}

A more detailed view at the properties of the outflows can be obtained from Figs~\ref{fig:M02D}-\ref{fig:M10k2D}. There, we visualize the distribution of the outflows in 2D planes : $Y_e$ vs $v_\infty$; $Y_e$ vs specific entropy; and $Y_e$ vs $\cos{\theta}$, with $\theta$ the angle between the polar axis and the velocity of the outflows. We first note that for all systems there is a clear correlation between electron fraction and entropy for $Y_e \lesssim 0.3$. Hotter material interacts more strongly with neutrinos, driving it to a less neutron-rich composition. This relationship appears to break at $Y_e\sim 0.3$; by that point hotter material is predicted to gain more velocity as it expands (as our method to estimate $v_\infty$ assumes conversion of thermal energy into kinetic energy as the gas expands), but not to become significantly less neutron-rich. This may be due in part to the short time available for neutrino-matter interactions, and in part to the fact that unbound matter is subjected to a higher flux of $\bar \nu_e$ than $\nu_e$.

Beyond that correlation, we see that the geometry and thermodynamics of the outflows are very variable, even for the relatively modest changes of binary parameters considered here. For M140-M130, the dominant hot outflows have a broad angular and velocity distribution, while the cold neutron-rich tidal ejecta is naturally confined to the equator. For M136-M126, the hot outflows dominate the mass budget, with otherwise a fairly similar distribution of velocity and entropy. For M127-M118, on the other hand, the hotter / less neutron rich outflows are confined to the polar regions  ($\cos\theta > 0.75$). The dominant outflow component at $Y_e\sim 0.2$ and $s\sim 15k_B$ per baryon has a broad angular distribution that nonetheless avoid the poles, and is slower than the polar ejecta (M136-M126 shows similar but weaker correlations between composition and orientation/velocity; M140-M130 does not). The cold tidal ejecta remains confined to the equator. The different components of the ejecta in M127-M118Adapt and M127-M118Ker have very similar properties, with more $Y_e\sim 0.2$ ejecta for the latter.

These variations illustrate the difficulty inherent in building simple semi-analytical models for the outflows. One division that seems consistent between all simulations presented here is between the hot polar outflows, near isotropic warm/hot outflows, and equatorial cold outflows, each with their own composition and entropy -- though the division between the first two may be somewhat arbitrary in some systems (e.g. M140-M130). It is also clear that neutrino physics that has been so far ignored in merger simulations has a noticeable impact on early-time outflows in neutron star mergers -- though more through the total amount of mass ejected by the system than the composition/velocity of each outflow component.

\subsection{Neutrino physics}

In~\cite{Rath:2026vfr}, we already used an early snapshot of M127-M118Adapt to show the expected impact of using the simulated neutrino spectrum in the calculation of $\nu\bar\nu \leftrightarrow e^+e^-$ rates, as well as the addition of inelastic scattering. We found that inelastic scattering was likely to be important for heavy-lepton neutrinos, who then remain thermally coupled to the fluid down to lower densities. Calculations of the pair annihilation rate using the simulated distribution of antineutrinos (rather than an equilibrium distribution), on the other hand, significantly increased pair annihilation rates in low-density regions. We now revisit these results with a simulation in which these reactions 
are self-consistently included in the evolution of the binary (M127-M118Ker).

\subsubsection{Neutrino packet weights and accuracy of the distribution function}

Before a more detailed analysis of neutrinos in our simulations, it is worth noting that an objective of these simulations is to try to use packets that each represent a change $\Delta f_{\nu} < 1$ in the distribution function $f_\nu$ of neutrinos, after integration over the direction of propagation of the neutrinos. In our simulations, $\Delta f_\nu =  \xi_s \Delta f_0$ with $\Delta f_0=10^{-3}$ and $\xi_s=1$ initially but growing over time to keep the total number of packets per species below $\sim 10^8$. In simulation M127-M118Adapt, we have $\Delta f_\nu = (0.03, 0.13, 0.07)$ for $(\nu_e, \bar\nu_e, \nu_x)$ right after merger; $\Delta f_\nu = (0.09, 0.22, 0.08)$ after $5\,{\rm ms}$; and $\Delta f_\nu = (0.16, 0.50, 0.27)$ after $10\,{\rm ms}$. The main reason behind the growth of $\Delta f_\nu$ is the rapid increase in the temperature of the remnant as we approach collapse to a black hole, which is associated with an increase in the total energy contained in neutrinos in the simulation domain: $E_\nu = (0.0002, 0.004, 0.034)M_\odot c^2$ at those same times. The adaptive weighting scheme does a reasonable job in limiting the growth in $\Delta f_\nu$ as $E_\nu$ increases, but $\Delta f_\nu$ still increases by a factor of $4$ for $\bar \nu_e$ as $E_\nu$ growth by two orders of magnitude. For M127-M118Ker, which reaches higher temperatures, the final values are $\Delta f_\nu = (0.22,0.62,0.37)$. By the end of these simulations, we are clearly stretching the limit of what $\Delta f_\nu$ can be while still remaining useful to calculate interaction rates. In particular, having a large $\Delta f_\nu$ is relatively safe for the calculation of the annihilation rate $\nu\bar\nu\rightarrow e^+ e^-$ and the `emission' component of inelastic scattering rates, because in practice the energy deposition from these reactions is averaged over many time steps. On the other hand, a large $\Delta f_\nu$ is a significant problem when calculating blocking factors. Nevertheless, these results are very encouraging. In~\cite{Foucart:2025nub}, we estimated that getting $\Delta f_\nu\sim 0.1$ was the best case scenario without increasing the number of Monte Carlo packets -- and that requires weighting schemes that make the dense regions unstable in our simulations. We see here that getting to $\Delta f_\nu\sim 0.1$ even just before collapse of the remnant to a black hole would be feasible with an increase in the number of packets of only a factor of $\sim 5$ -- possibly less if we force the same $\Delta f_\nu$ for all species. Even the higher values of $\Delta f_\nu$ used here allow for evolutions in which reaction rates are directly calculated from $f_\nu$, at the cost of increased shot noise in the Monte Carlo code. 

We note that there is a very sigificant difference between what is needed to calculate reaction rates from the simulated distribution of neutrinos, and what is needed to accurately measure $f_\nu$ within a cell. This is because in many regions of phase space, $f_\nu \ll 1$. When $f_\nu \ll \Delta f_\nu \ll 1$, we clearly cannot measure $f_\nu$, but that does not stop us from calculating emission or absorption rates. Factors of $f_\nu$ and $(1-f_\nu)$ in the reaction rates will average to the correct values over time, as long as enough packets are sampled over the timescale for the thermodynamical evolution of the remnant and $(1-f_\nu)>0$. Increasing $\Delta f_\nu$ to a point where sampling noise in the simulation regularly creates regions of phase space where we measure $f_\nu>1$ is what will create biases in the average reaction rates used in the simulation -- and when $\Delta f_\nu > 0.5$, this clearly becomes a concern. Overall, simulations M127-M118Adapt and M127-M118Ker should thus be seen as a first attempt at directly using $f_\nu$ in simulations, with more efforts definitely required to get to a point where interaction rates can be computed to high-accuracy.

\subsubsection{Representative simulation snapshot}

\begin{figure*}
 \includegraphics[width=0.95\textwidth]{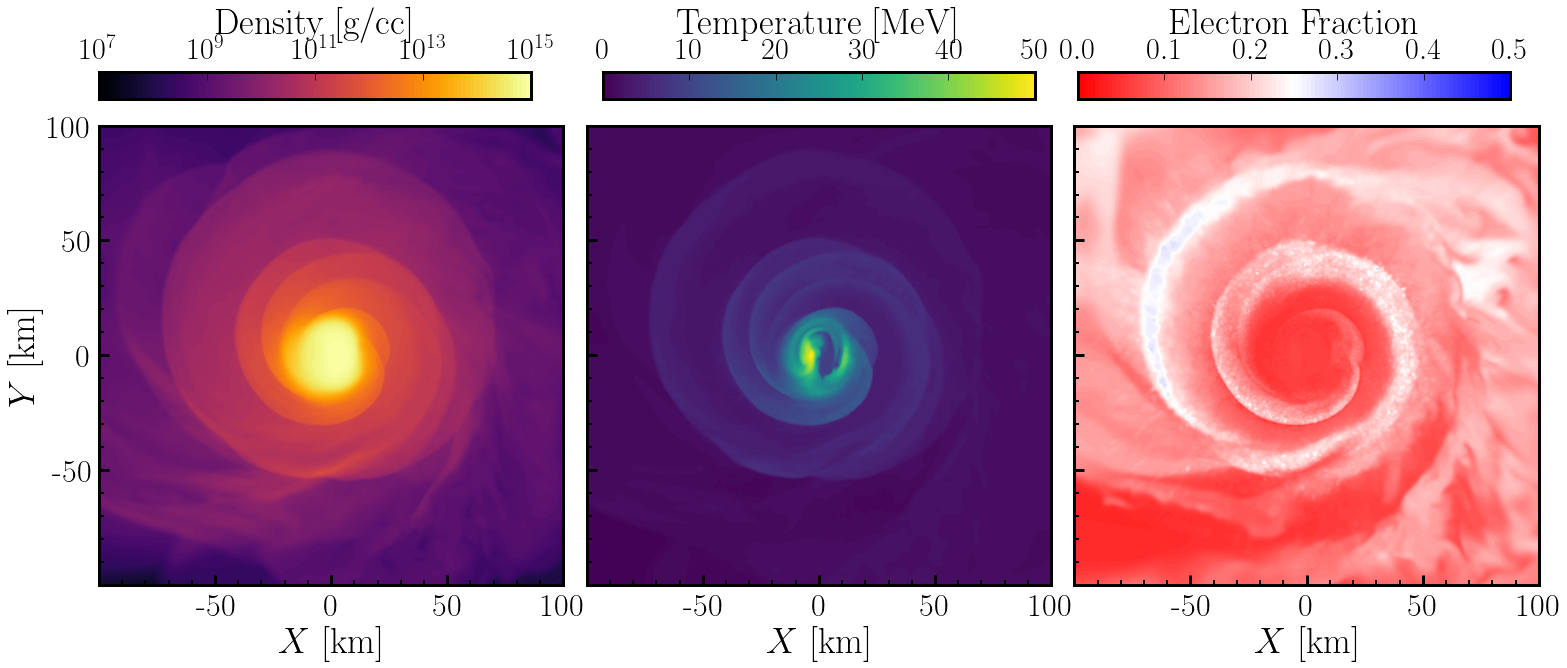}
 \caption{Baryon density (left), temperature (center) and electron fraction (right) within the orbital plane of simulation M127-M118Adapt $5\,{\rm ms}$ post-merger.}
\label{fig:HorSlice}
\end{figure*}

\begin{figure*}
\includegraphics[width=0.95\textwidth]{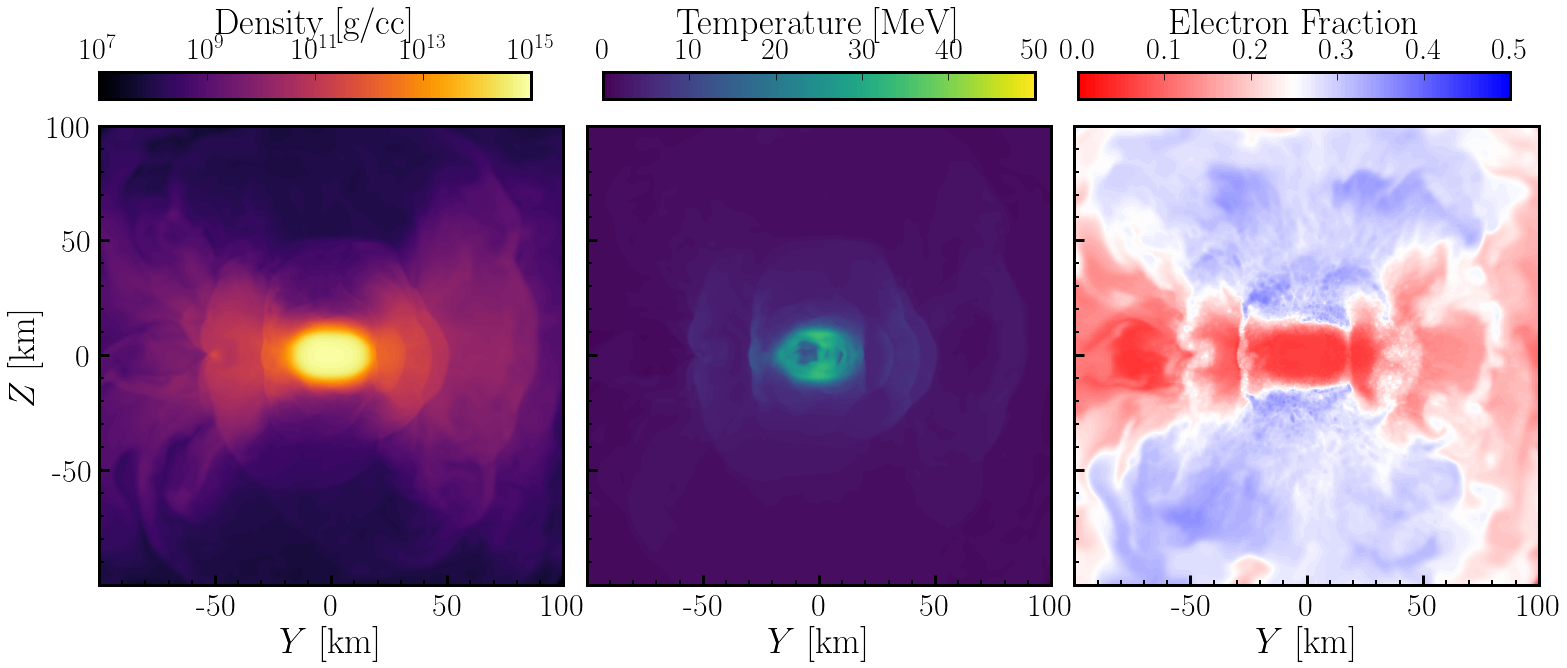}
 \caption{Same as Fig~\ref{fig:HorSlice}, but for a slice orthogonal to the orbital plane ($x=0$ in grid coordinates).}
\label{fig:VerSlice}
\end{figure*}

\begin{figure*}
\includegraphics[width=0.95\textwidth]{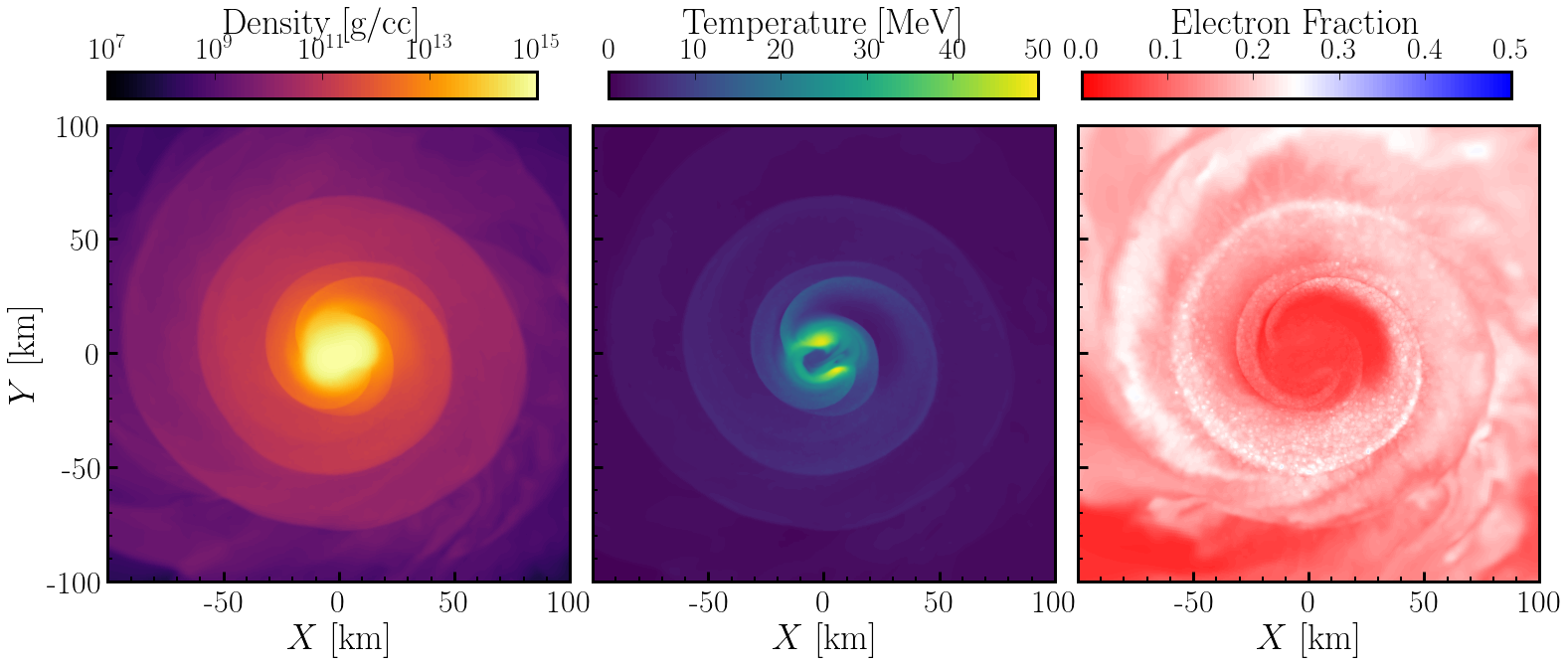}
 \caption{Same as Fig~\ref{fig:HorSlice}, but simulation M127-M118Ker.}
\label{fig:HorSlice2}
\end{figure*}

We start our disucussion of neutrinos and neutrino-matter interactions by analyzing a snapshot of simulations  M127-M118Adapt and M127-M118Ker, $5\,{\rm ms}$ after merger. The density, temperature and electron fraction of the system are visualized in two orthogonal slices in Figs~\ref{fig:HorSlice}-\ref{fig:VerSlice}. We see the dense, neutron rich remnant with a hot surface and the slightly less neutron-rich accretion disk within which shocked tidal arms are clearly visible. Lower density outflows are driven from the disk and remnant, and these outflows have $Y_e>0.25$ in the polar regions. Shot noise from the coupling of the Monte Carlo transport with the fluid evolution is visible when looking at $Y_e$. Qualitatively, all binary neutron star mergers performed with neutrino transport show similar features before collapse of the remnant to a black hole (except for the shot noise, which is specific to Monte Carlo transport). Fig.~\ref{fig:HorSlice2} shows simulation M127-M118Ker at a similar time. The same qualitative features are present, but we note small quantitative differences in the temperature profile (hotter hot spot), composition profile, and structure of the tidal arms. Vertical slices through M127-M118Ker are harder to directly compare to M127-M118Adapt as the two simulations are not exactly in phase and we do not have snapshots at exactly the same time -- other slices do not show any surprising features, however.

\begin{figure*}
\includegraphics[width=0.95\textwidth]{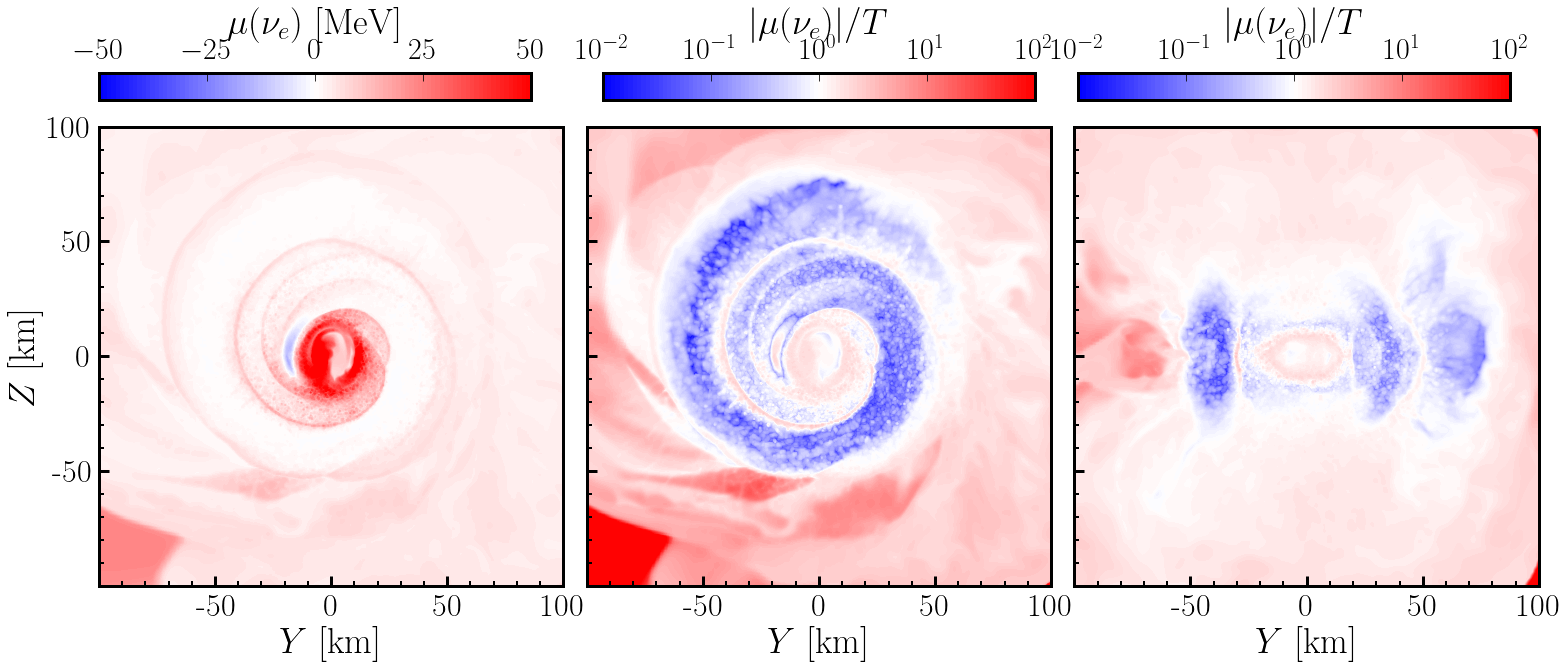}
 \caption{{\it Left}: Equilibrium neutrino checmical potential $\mu_{\nu_e} = \mu_e + \mu_p - \mu_n$ on the same equatorial slice as Fig.~\ref{fig:HorSlice}. {\it Center}: $|\mu_{\nu_e}/T|$ on the same slice. {\it Right}: $|\mu_{\nu_e}/T|$ on the same vertical slice as in Fig.~\ref{fig:VerSlice}. Note the log scale on the central and rightmost panels; blue regions in those panels are close to their equilibrium $Y_e$.}
\label{fig:munu}
\end{figure*}

In Fig.~\ref{fig:munu}, we show $\mu_e + \mu_p - \mu_n$, with $\mu_{e,p,n}$ the chemical potentials of electrons, protons and neutrons. In neutrinoless beta-equilibrium, this quantity would vanish. In most regions of the remnant, however, we do not reach an equilibrium composition. In the densest regions, the neutron star is more neutron-rich than required for neutrinoless beta equilibrium, due to its increased temperature. Neutrinos quickly equilibrate with the fluid at fixed electron lepton number, but diffuse out too slowly for the system to reach $\mu_e + \mu_p - \mu_n=0$. The chemical potential of electron neutrinos in equilibrium with the fluid is then $\mu_{\nu_e} = \mu_e + \mu_p - \mu_n$. In the hottest regions of the core, we find $\mu_{\nu_e} \ll -T$. The production of $\nu_e$ is then suppressed while the production of $\bar\nu_e$ is enhanced. In the disk, the fluid reaches its equilibrium composition and $|\mu_{\nu_e}| \ll T$. In the outflows, the matter is again more neutron-rich than what equilibrium considerations would suggest. As already discussed, this is in part because of the large velocity of the outflows and in part due to irradiation by (mostly) $\bar\nu_e$. M127-M118Ker and M127-M118Adapt are very similar in that respect.

\begin{figure*}
\includegraphics[width=0.95\textwidth]{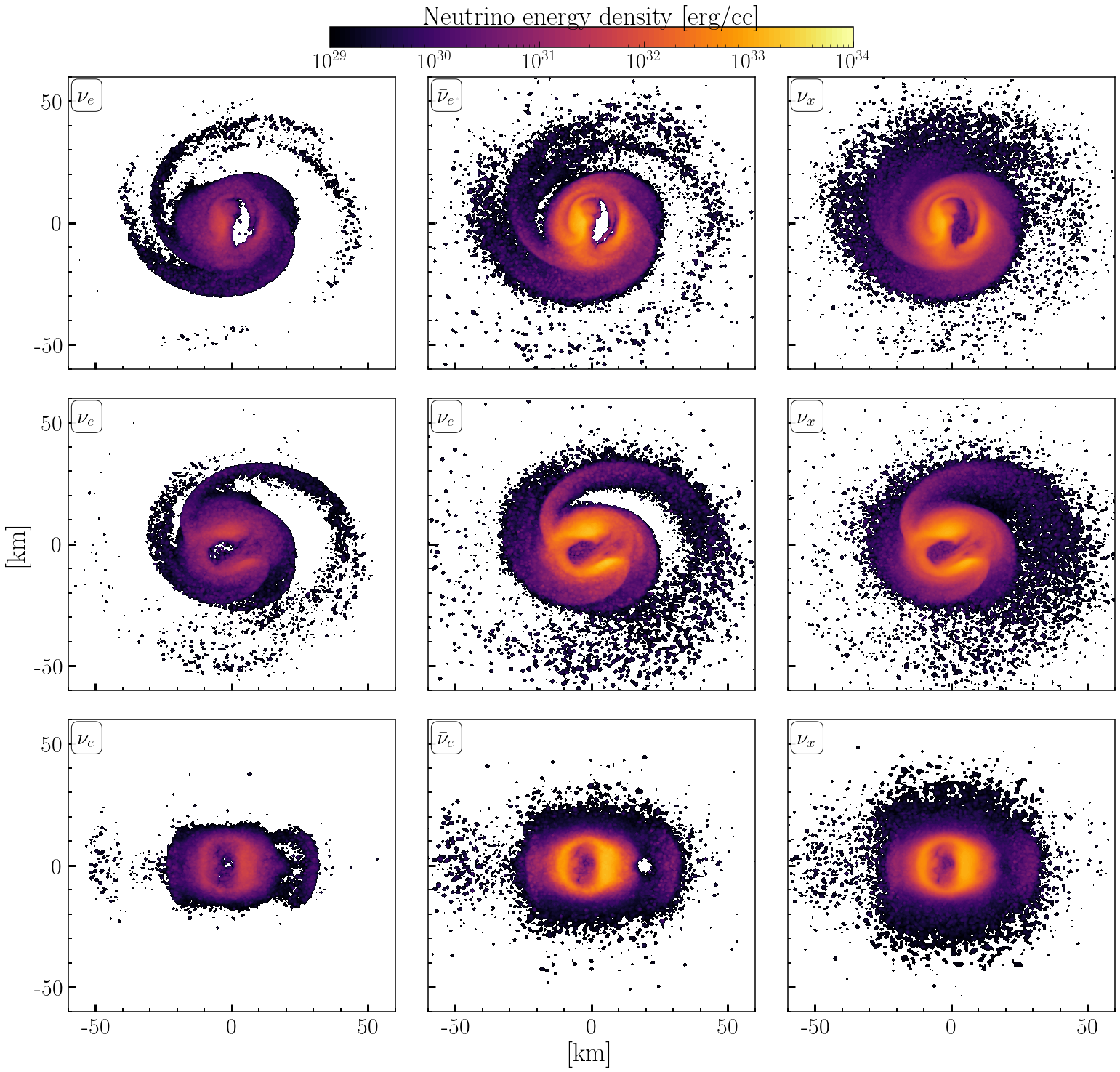}
 \caption{Energy density of neutrinos on the same equatorial slice of M127-M118Adapt as in Fig.~\ref{fig:HorSlice} ({\it Top}); on the same equatorial slice for simulation M127-M118Ker ({\it Middle}), and on a vertical slice of M127-M118Ker ({\it Bottom}). Note that we zoomed in more on the remnant than in previous figures.  We show results for $\nu_e$ ({\it Left}),  $\bar\nu_e$ ({\it Center}), and  $\nu_x$ ({\it Right}, all four species combined). For reference, the maximum rest mass energy density on the grid is $\sim 0.002$ in these units.}
\label{fig:Enu}
\end{figure*}

The resulting spatial distribution of neutrinos is shown on Fig.~\ref{fig:Enu}. We see that even on a log scale, neutrinos are heavily concentrated in the hottest regions of the remnant and immediately outside the neutron star. As expected from the chemical potentials, the density of $\bar \nu_e$ is much larger than the density of $\nu_e$. Heavy lepton neutrinos are dominant in some regions of the remnant disk and in low-density polar regions -- though this is largely because we combine all 4 species of heavy lepton neutrinos together in those figures. Besides differences due to the temperature and density profile of the tidal arms, the main notable difference between M127-M118Ker and M127-M118Adapt is more trapping of the heavy-lepton neutrinos when using kernels. With the use of kernels, the distribution of $\nu_x$ is much closer to the distribution of $\bar\nu_e$ than without kernels. This is an expected effect of the inclusion of inelastic scattering, though one that could not be observed in earlier neutron star merger simulations. A secondary difference is higher neutrino energy densities in cold, dense regions for $\nu_e$ and $\bar\nu_e$ when using kernels.

\subsubsection{Opacities and energy distribution of neutrinos}

\begin{figure*}
\includegraphics[width=0.95\textwidth]{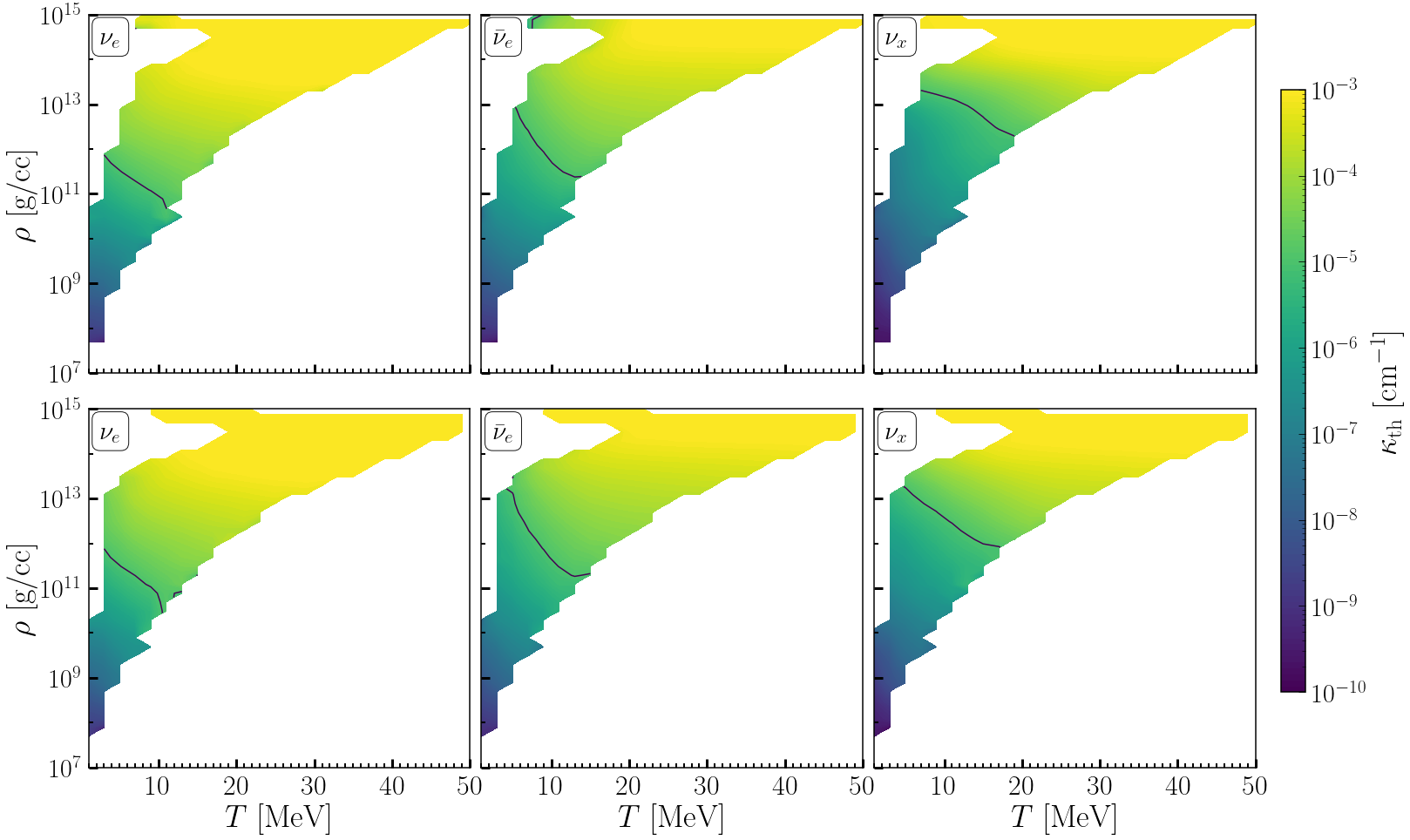}
 \caption{Average thermalization opacity $5\,{\rm ms}$ after merger in simulation M127-M118Adapt ({\it Top}) and M127-M118Kernel ({\it Bottom})  for $\nu_e$ ({\it Left}),  $\bar\nu_e$ ({\it Center}), and  $\nu_x$ ({\it Right}. The black line shows the $\kappa_{\rm th}=1\,{\rm km^{-1}}$ contour. The average is weighted by the energy of neutrinos within each density-temperature bin. White regions correspond to values of the density and temperature that do not occur within these snapshots.}
\label{fig:kth}
\end{figure*}

To visualize more easily changes to the opacity and energy distribution of neutrinos under different thermodynamical conditions, we now consider a different set of visualizations. We divide the computational domain into $50\times 50$ bins based on the fluid density and temperature. The density bins are logarithmically spaced in $[6.2\times 10^8,6.2\times 10^{15}]\,{\rm g/cm^3}$. The temperature bins are linearly spaced in $[0,50]\,{\rm MeV}$. We first look in Fig.~\ref{fig:kth} at the average thermalization opacity, with the average weighted by the energy of the Monte Carlo packets within each bin. We approximate the thermalization opacity as
\beq
\kappa_{\rm th} = \sqrt{(\kappa_a + \kappa_{\rm inel})(\kappa_a + \kappa_{\rm inel} + \kappa_{\rm el})}.
\eeq
Here, $\kappa_a$ is the opacity due to all absorption processes (including pair annihilation, if included in the simulation); $\kappa_{\rm el}$ is the opacity to elastic scattering (usually written $\kappa_s$), and $\kappa_{\rm inel}$ is the opacity to inelastic scattering on electrons. We note that this is an approximation, as this formula implicitly assumes that inelastic scattering is as efficient as absorption/emission at thermalizing neutrinos. This is the same method used to calculate the thermalization opacity as in~\cite{Rath:2026vfr}, where we already explored the impact of using kernels to calculate reaction rates, as well as the difference between using energy-dependent opacities and `gray' opacities assuming neutrinos in thermal equilibrium with the fluid. The main difference here is that simulation M127-M118Ker consistently used kernels for the calculation of emissivities and opacities during the entire simulation. The results are nevertheless consistent with~\cite{Rath:2026vfr}: the most notable difference is the increased opacity to $\nu_x$ around and outside of the decoupling region (close to the $\kappa_{\rm th}=1\,{\rm km^{-1}}$ contour). We also note an increased opacity for $\nu_e, \bar\nu_e$ in cold, dense regions where the charged current reactions considered here are suppressed. As in Fig.~\ref{fig:Enu}, we note that heavy lepton neutrinos are much more similar to $\bar \nu_e$ in the M127-M118Ker simulation than in M127-M118Adapt.

\begin{figure}
\includegraphics[width=0.9\columnwidth]{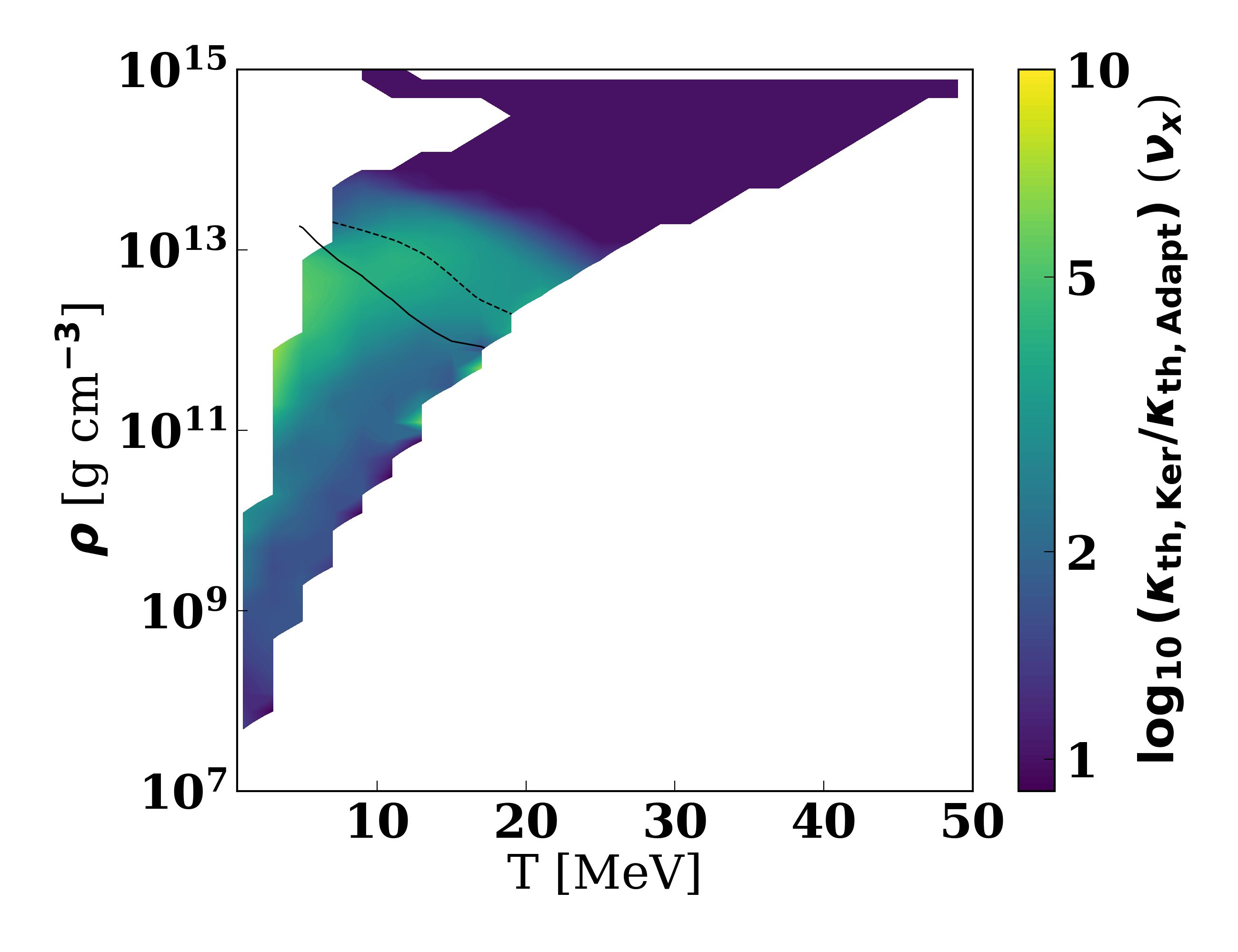}
 \caption{Ratio of the thermalization opacity for $\nu_x$ between simulation M127-M118Ker and M127-M118Adapt, $5\,{\rm ms}$ after merger. The solid and dashed line show the $\kappa_{\rm th}=1\,{\rm km^{-1}}$ contour for M127-M118Ker and M127-M118Adapt, respectively.}
\label{fig:kratio}
\end{figure}

Outside of those cold dense regions, $\kappa_{\rm th}$ for $\nu_e$ and $\bar\nu_e$ is consistent between the two simulations at the $(10-20)\%$ level. For $\nu_x$, the ratio between the opacity in the two simulations is shown on Fig.~\ref{fig:kratio}. We see that outside the decoupling region, the opacity in M127-M118Ker is $2-5$ times higher than in M127-M118Adapt. We note in particular that the absorption rate of $\nu_x$ is higher for M127-M118Ker in the dense, hot tidal arms that are known to be important to the generation of outflows (see e.g.~\cite{Radice:2023xxn}). Parts of the tidal arms at density $\rho\sim 10^{12}\,{\rm g/cm^3}$ will see enhanced energy transfer from the hot shocked regions to the lower density / colder regions that surround them. This may partially explain the increased mass ejection in M127-M118Ker.

\begin{figure*}
\includegraphics[width=0.97\textwidth]{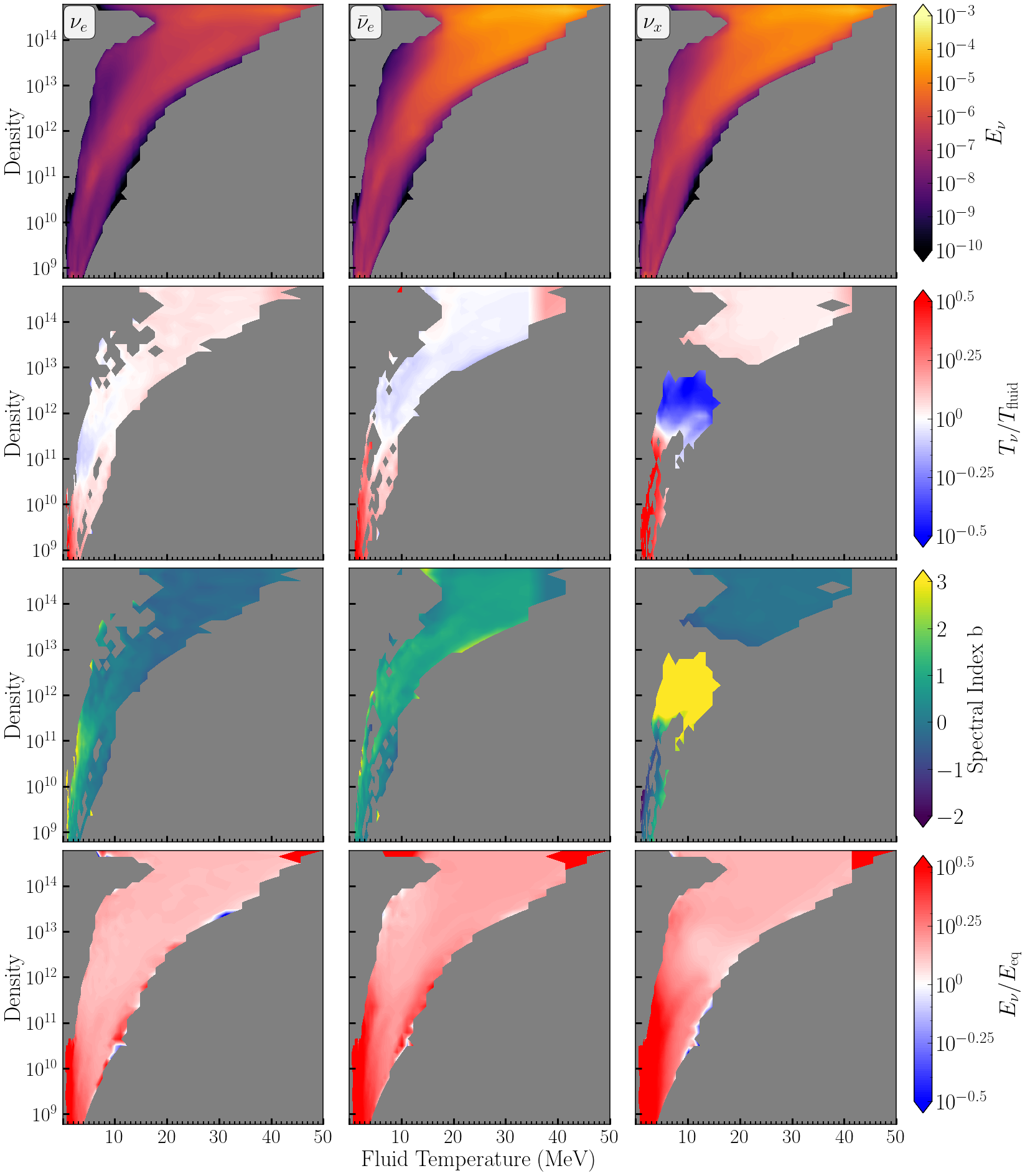}
 \caption{Neutrino properties $5\,{\rm ms}$ after merger for simulation M127-M118Adapt. {\it Top}: Neutrino energy within each bin, in $G=c=M_\odot=1$ units. {\it Second row}: Neutrino temperature $T_\nu$, normalized to the fluid temperature. {\it Third row}: Spectral index $b$, as defined in the text. {\it Bottom}: Average energy of neutrinos in each bin, normalized to the expected equilibrium energy for a vanishing neutrino chemical potential. We show results for $\nu_e$ ({\it Left}),  $\bar\nu_e$ ({\it Center}), and  $\nu_x$ ({\it Right}).}
\label{fig:spectraA}
\end{figure*}

\begin{figure*}
\includegraphics[width=0.97\textwidth]{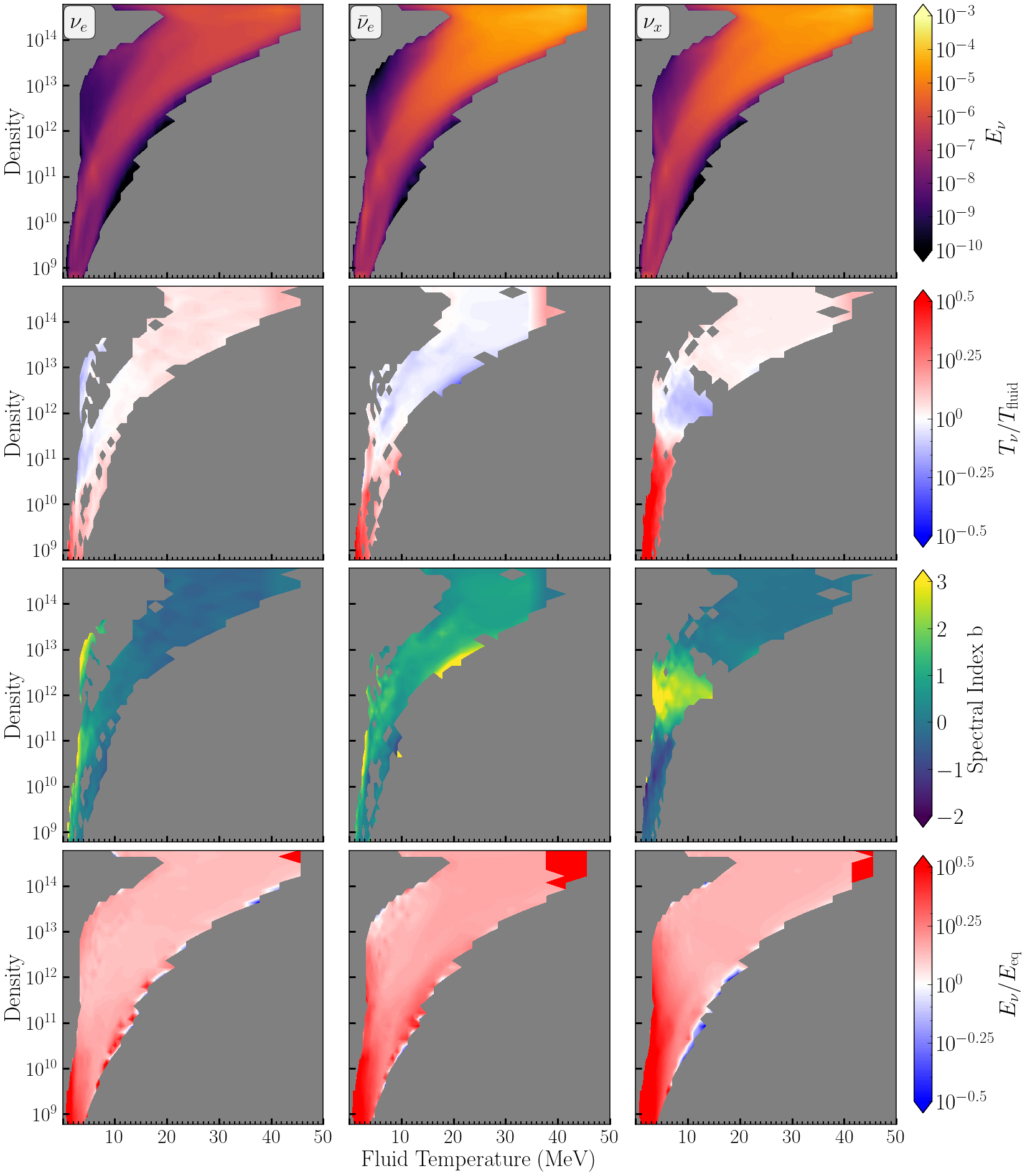}
 \caption{Same as Fig.~\ref{fig:spectraA}, but for simulation M127-M118Ker.}
\label{fig:spectraK}
\end{figure*}

To analyze the impact of the modified set of reactions on the neutrino distribution function, we model the energy spectrum of all neutrinos contained within one of our density-temperature bins. We have $50^2$ bins and $10^8$ Monte Carlo packets per species over the whole simulation domain -- thus in regions of the density-temperature plane that exist in our simulations, most bins will have enough packets to capture the shape of the neutrino energy distribution. To visualize that distribution across the entire parameter space of the simulation, we model the distribution function of neutrinos as
\beq
f_\nu (\epsilon)= A\frac{\epsilon^b}{1+\exp{(\epsilon/T_\nu})}
\eeq
with $A, b, T_\nu$ free parameters, and $\epsilon$ the energy of individual neutrinos. In practice, we bin neutrinos into the same $16$ energy bins used in our NuLib table for interaction rates (measured in the fluid frame), and calculate the total energy within each bin $E_i$. We then fit those 16 data points to the expected value
\beq
E_{i,exp} = \tilde A \int_{\epsilon_i}^{\epsilon_{i+1}} \frac{\epsilon^{3+b}}{1+\exp{(\epsilon/T_\nu)}}
\eeq
with $\epsilon_i$ the energy of the bottom of the $i^{\rm th}$ bin and $\tilde A$ a constant proportional to $A$. We note that our model for $f_\nu$ is different from the Fermi-Dirac distribution
\beq
f^{\rm FD}_\nu (\epsilon)= \frac{1}{1+\exp{[(\epsilon-\mu)/T_\nu}]}
\eeq
with $\mu$ the neutrino chemical potential. Our choice reflects our desire to capture the shape of the neutrino energy distribution in out-of-equilibrium regions. The free parameters of the fit are $\tilde A, b, T_{\nu}$. Adding the chemical potential as a fourth parameter leads to large degeneracies between fitting parameters, as might be expected considering that we are fitting 16 data points, and that in addition high-energy bins are often empty. In Figs.~\ref{fig:spectraA}-\ref{fig:spectraK}, we show results for simulations M127-M118Adapt and M127-M118Kernel. We only show the fit parameters when the $1-\sigma$ error from the fit is below $0.3T_\nu$ for the temperature and below $0.15 (1.0+|b|)$ for the spectral index. We note that the main source of uncertainty in the fit here is the unknown functional form of the neutrino energy distribution, not the number of Monte Carlo packets available. In regions where we do not provide best-fit parameters, our functional form either does not fit well the simulation data, or has large degeneracies between $T_\nu$ and $b$. It is also worth emphasizing that as all grid cells with similar densities and temperatures are binned together, this is a fit to an average distribution function in that region of parameter space, ignoring spatial variations and variations with the fluid composition (although in most bins, the spread in the sampled value of the electron fraction is small~\cite{Rath:2026vfr}).

\begin{figure}
\includegraphics[width=0.9\columnwidth]{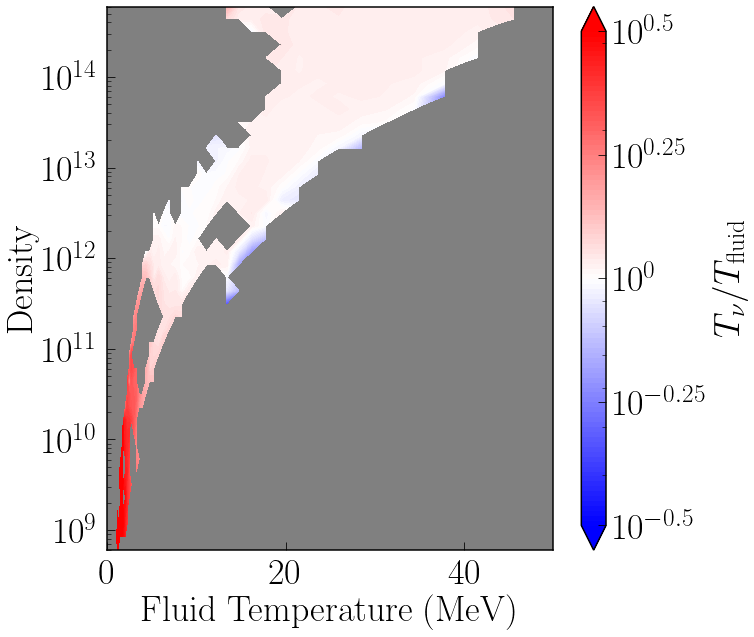}\\
\includegraphics[width=0.9\columnwidth]{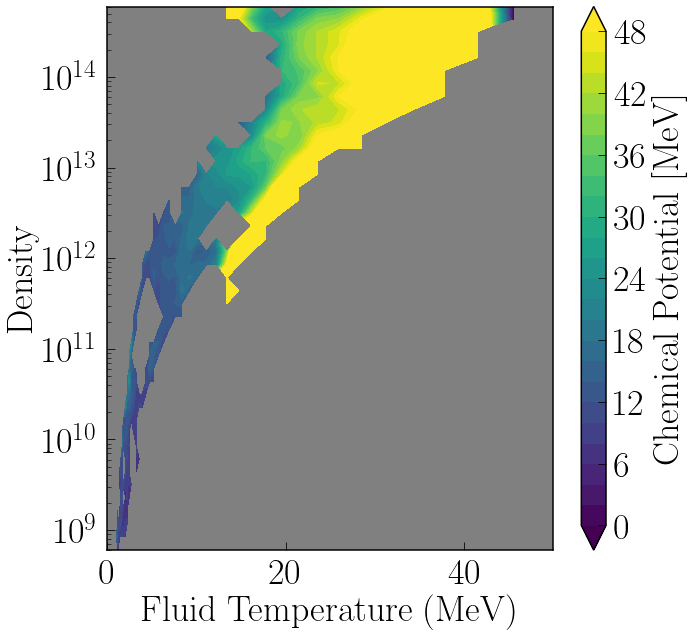}
 \caption{Fitting to a rescaled Fermi-Diract distribution function for $\bar\nu_e$ in simulation M127-M118Adapt. {\it Top}: neutrino temperature; {\it Bottom} Chemical potential of $\bar\nu_e$.}
\label{fig:spectraMuA}
\end{figure}

In each figure, the top row shows where neutrinos can be found within the simulation, for reference. We then show the best-fits $T_\nu/T_{\rm fluid}$ and $b$. For $\nu_e$, we find values close to the $\mu_\nu=0$ weak equilibrium in dense regions, and a harder spectrum $b\sim 2$ in low-density regions. For $\bar\nu_e$, the dense regions have a harder spectrum and lower temperature than the $\mu_\nu=0$ weak equilibrium predictions. We aso show the average neutrino energy, which does not require fitting any spectrum. The average energy is slightly higher than the weak equilibrium prediction. These differences for $\bar\nu_e$ at high density are due to the large chemical potential $\mu_{\bar \nu_e}$ observed there. We confirm this in Fig.~\ref{fig:spectraMuA}, where we fit to a Fermi-Dirac distribution function instead (with free parameters $T_\nu,\mu$, and an overall scaling factor) and find $T_{\nu}\approx T_{\rm fluid}$ and large positive chemical potentials for $\bar\nu_e$. Because each density-temperature bin contains a range of electron fractions, we cannot directly compare the results of the fit to a single value of the neutrino chemical potential in the simulation. Qualitatively, however, the large potentials returned by the fit in dense, hot regions are consistent with our simulation results (see Fig.~\ref{fig:munu}). Unfortunately, for $\nu_e$ and $\nu_x$, using the Fermi-Dirac distribution either fails to model the energy spectrum or leads to large degeneracies between the fitting parameters in large parts of the parameter space. Even for $\bar\nu_e$, it fails to provide a good fit more often than our original ansatz in the low-density regions. For $\nu_e$, however, using a Fermi-Dirac distribution in which we force $\mu_\nu=0$ provides a good fit to the data, with $T_{\nu}\approx T_{\rm fluid}$. This is likely because a spectrum with a large negative neutrino potential is very similar to a rescaled spectrum with $\mu_\nu=0$ (i.e. up to a global rescaling by $e^{(-\mu/T)}$), except at low neutrino energies. There is thus a degeneracy between $\mu_\nu$ and the rescaling factor when $\mu_\nu \ll T$.

For heavy-lepton neutrinos, we see more significant differences between M127-M118Adapt and M127-M118Ker. In the region $10^{11}\,{\rm g/cm^3} < \rho < 10^{13}\,{\rm g/cm^3}$, where most neutrino-matter interactions involving $\nu_x$ are scattering events and not emissions/absorptions, the energy distribution of $\nu_x$ is very non-thermal -- even though the average energy of neutrinos is close to its equilibrium value. Specifically, our best fit values imply $T_{\nu}\ll T_{\rm fluid}$ and $b \gg 0$ for M127-M118Adapt. For M127-M118Ker, inelastic scatterings on electrons bring the energy distribution closer to thermal equilibrium, but the deviations remain more significant than for electron type neutrinos. We also note that around $10^{13}\,{\rm g/cm^3}$, our chosen functional form does not fit the simulated spectrum well for M127-M118Adapt, while it performs better for the more thermalized neutrinos from M127-M118Ker. Visualizing the spectrum in regions where strong non-thermal features are observed, we find that the energy distribution is more narrowly peaked than what a thermal spectrum would predict. This can be understood from two effects. First, low energy neutrinos are free to escape the system even when the average opacity is large. Second, inelastic scattering do not redistribute energy as efficiently as emission-absorption processes: each scattering event changes the energy of neutrinos by a fraction of the fluid frame energy of the target electron, rather than the energy of the new neutrino being drawed from the full emission spectrum. 

Modeling the energy spectrum of neutrinos is an important component in the calculation of interaction rates in energy-integrated schemes. From our simulations, we can gather a few important pieces of information. First, in dense regions, the spectrum of $\bar\nu_e$ is well modeled by a Fermi-Dirac distribution with a positive chemical potential. The spetrum of $\nu_e$ is well modeled by a Fermi-Dirac distribution with vanishing chemical potential. In lower density regions, both species are better modeled by a hard spectrum ($b>0$) with $T_\nu >T_{\rm fluid}$. For $\nu_x$, the region where absorption opacities are low but scattering opacities are higher are best modeled by a hard spectrum that roughly match the average neutrino energy of the equilibrium spectrum. In lower density regions, we find a soft spectrum ($b<0$) at $T_\nu >T_{\rm fluid}$.

\subsubsection{Neutrino luminosities and spectra}

\begin{figure}
\includegraphics[width=0.9\columnwidth]{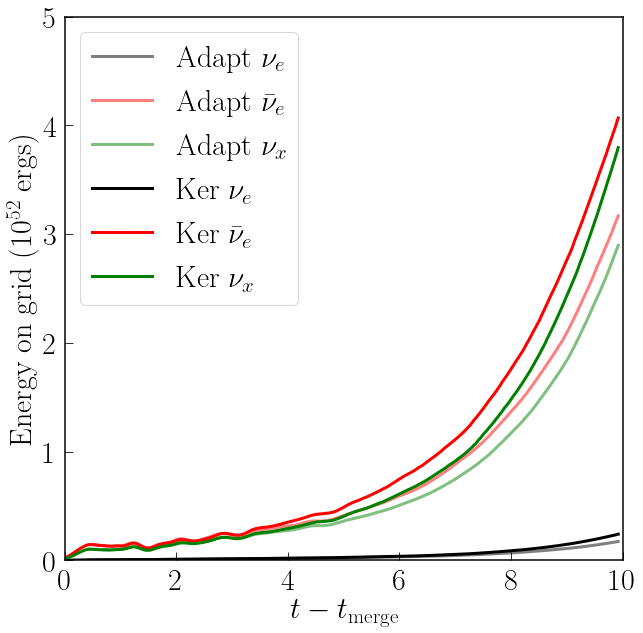}
 \caption{Total energy of the neutrinos on the computational grid, measured in the inertial frame. We show results for M127-M118Adapt and M127-M118Ker and for all 3 evolved neutrino species. Results for $\nu_x$ are the sum of all heavy-lepton neutrinos.}
\label{fig:EnNu}
\end{figure}

\begin{figure}
\includegraphics[width=0.9\columnwidth]{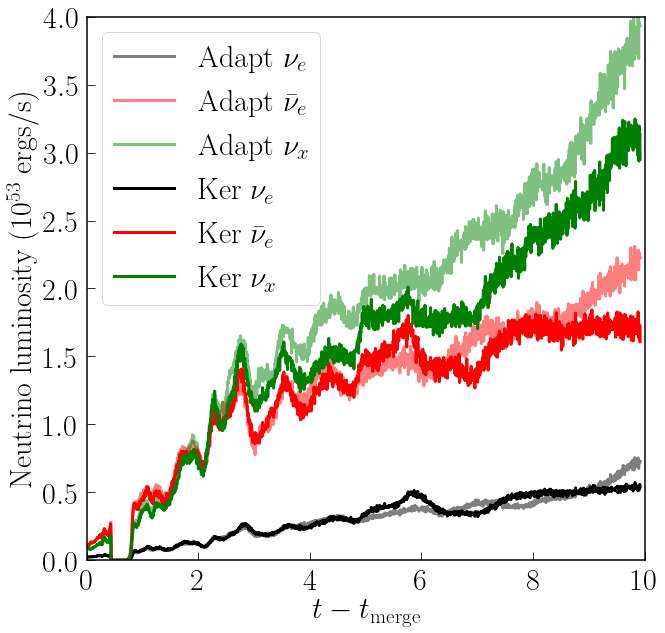}
 \caption{Luminosity of neutrinos leaving the computational grid, measured in the inertial frame. We show results for M127-M118Adapt and M127-M118Ker and for all 3 evolved neutrino species. Results for $\nu_x$ are the sum of all heavy-lepton neutrinos. The dip just before $1\,{\rm ms}$ corresponds to a change from a grid just surrounding the neutron stars to the larger grid used in post-merger evolution.}
\label{fig:LumNu}
\end{figure}

We now turn to more global properties of the neutrinos, still focusing on our comparison between M127-M118Adapt and M127-M118Ker. Fig.~\ref{fig:EnNu} shows the total energy of the neutrinos on the grid. We see that the energy budget is about equally divided between $\bar\nu_e$ and $\nu_x$. The energy in $\nu_e$ is more than an order of magnitude smaller. As $\nu_x$ groups all heavy lepton neutrinos, each species has a total energy of about $25\%$ of the $\bar\nu_e$ energy. The total energy grows with the temperature of the remnant, and as regions with low emissivity but high scattering opacity get filled with diffusing neutrinos. Fig.~\ref{fig:LumNu} shows the neutrino luminosity for the same simulations. As the opacity of the remnant to $\nu_x$ is smaller than its opacity to $\bar\nu_e$, the former is now the dominant emission channel (on an aggregate basis -- the luminosity per species is higher for $\bar\nu_e$). More importantly, we see the clear impact of the modified set of reactions used in M127-M118Ker. Even though the remnant is slightly hotter in that simulation, the luminosity in $\nu_x$ is smaller. This is in part due to increased absorption rate for $\nu\bar\nu$ pair annihilation, and in part due to a reduction in the average energy of the neutrinos (see below).

\begin{figure}
\includegraphics[width=0.9\columnwidth]{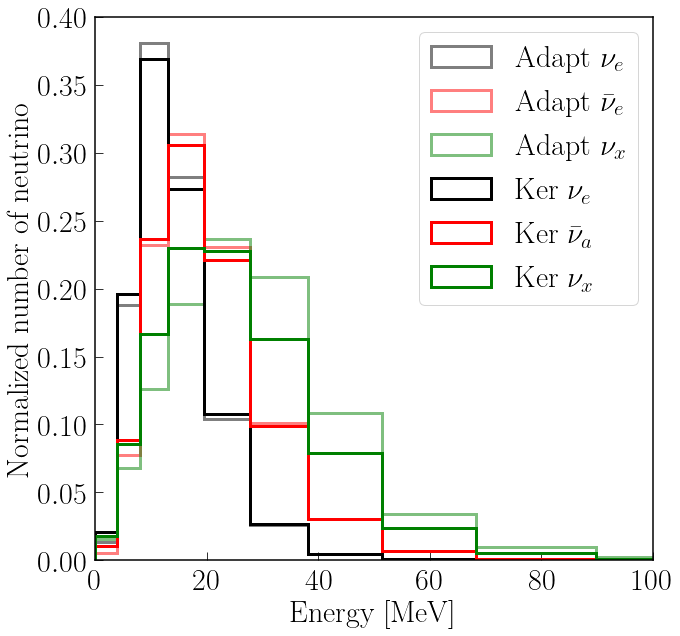}
 \caption{Spectrum of the neutrinos leaving the computational grid for M127-M118Adapt and M127-M118Ker and for all 3 evolved neutrino species. We show the number of neutrinos in each bin, normalized by the total number of neutrinos of that species.}
\label{fig:SpectraNu}
\end{figure}

\begin{table}
\begin{tabular}{c|ccccc}
Name & Species & $\langle \epsilon \rangle$ [MeV] &  $T_\nu$ [MeV] & $b$\\
\hline
M127-M118-Adapt & $\nu_e$ & $13.4$ & $2.61\pm 0.08$ & $1.77\pm 0.15$\\
M127-M118-Adapt & $\bar\nu_e$ & $18.8$ & $4.23\pm 0.10$ & $1.13\pm 0.09$\\
M127-M118-Adapt & $\nu_x$ & $25.4$ & $8.84\pm 0.47$ & $-0.26\pm 0.15$\\
\hline
M127-M118-Ker & $\nu_e$ & $13.3$ & $2.86\pm 0.08$ & $1.33\pm 0.12$\\
M127-M118-Ker & $\bar\nu_e$ & $18.4$ & $4.48\pm 0.09$ & $0.83\pm 0.08$\\
M127-M118-Ker & $\nu_x$ & $22.4$ & $8.01\pm 0.08$ & $-0.43\pm 0.03$\\
\end{tabular}
\caption{Summary of the properties of the neutrinos leaving the grid in simulations M127-M118-Adapt and M127-M118-Ker, up to $10\,{\rm ms}$ after merger. We show the average energy (weighted by the number of neutrinos), as well as the parameters $T_\nu$ (neutrino temperature) and $b$ (spectral index) from the best fit spectrum. Error bars are $1-\sigma$ errors obtained from the covariance matrix of the least-square fit.}
\label{tab:neutrinos}
\end{table}

\begin{figure*}
\includegraphics[width=0.95\textwidth]{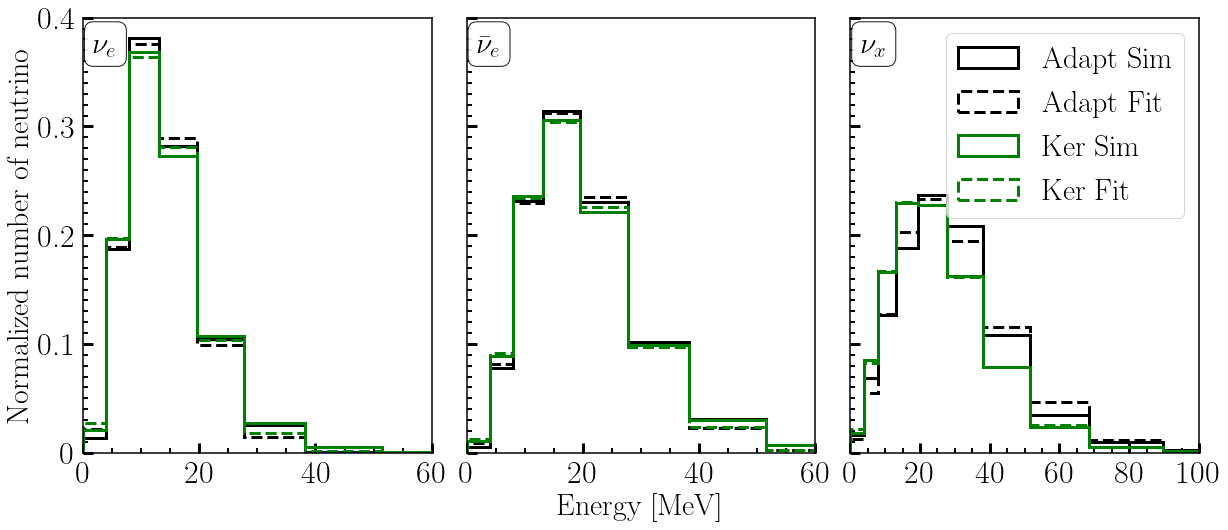}
 \caption{Same as Fig~\ref{fig:SpectraNu}, but now showing the simulated and best-fit spectra for $\nu_e$ ({\it Left}), $\bar\nu_e$ ({\it Middle}), and $\nu_x$ ({\it Right}).}
\label{fig:SpectraNuFit}
\end{figure*}

Fig.~\ref{fig:SpectraNu} shows the spectrum of neutrinos, plotting normalized neutrino numbers within each energy bin. We see that, as is usual in neutron star mergers, the energy of the neutrinos satisfy the hierarchy $\epsilon_{\nu_e}< \epsilon_{\bar\nu_e}<\epsilon_{\nu_x}$. The inclusion of inelastic scattering clearly brings down the energy of the heavy lepton neutrinos. To quantify these statements, we fit each spectrum using the ansatz
\beq
\frac{dN}{d\epsilon} = A\frac{\epsilon^{2+b}}{1+\exp{(\epsilon/T_\nu})},
\eeq
i.e. the same ansatz as for the local neutrino spectra (but for the neutrino number distribution instead of the energy distribution). For a black body at temperature $T$, we would expect $b=0$ and $T_\nu=T$. We find best fit values for $(A,T_\nu,b)$ 
using scipy's {\it curve\_fit} function. Results for $T_\nu$ and $b$ are shown in Table~\ref{tab:neutrinos}, together with the number-weighted average energy of each neutrino species. Inelastic scattering causes a $10\%$ decrease in the average energy of the neutrinos for $\nu_x$, and a slight softening of the spectrum for all species (decrease in $b$). Fig.~\ref{fig:SpectraNuFit} shows the quality of the fit in both simulations. For $\nu_x$, we see that our chosen ansatz fits the results from M127-M118-Ker, but is less accurate for M127-M118-Adapt. For $\nu_e$ and $\bar\nu_e$, we find similar errors in the fits from both simulations, and slight deviations from the chosen functional form. Most notably, the fits show a steeper decline of the spectra at high energy than the simulated data.

Finally, we consider variations in the energy of neutrinos with the angle $\theta$ between the rotation axis of the system and the direction of propagation of the neutrinos. The angular distribution of neutrinos is similar to what was found in previous Monte Carlo simulations, with a preference for emission in the polar direction by a factor of $2$ for $\nu_e$ and $\bar \nu_e$ and a factor of $3$ for $\nu_x$. The average energy of the neutrinos is higher in the polar regions. For $\nu_e$, that energy monotonously increases from $12.1\,{\rm MeV}$ to $14.5\,{\rm MeV}$.  For $\bar\nu_e$, it similarly varies from $17.1\,{\rm MeV}$ to $19.4\,{\rm MeV}$; and for $\nu_x$ from $20.2\,{\rm MeV}$ to $22.9\,{\rm MeV}$ (with all numbers reported for M127-M118-Ker).

\subsubsection{Pair annihilation in polar regions}

\begin{figure*}
\includegraphics[width=0.95\textwidth]{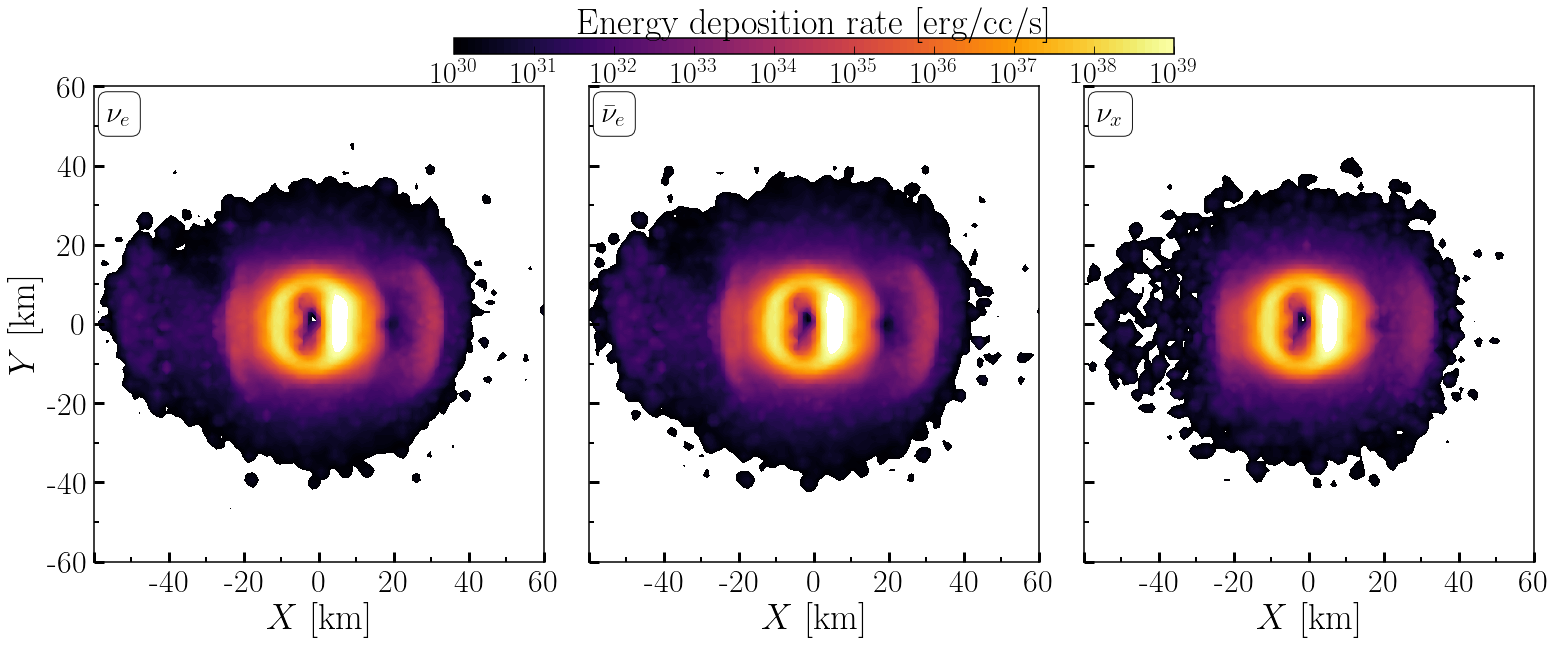}
 \caption{Energy deposition from $\nu\bar\nu\rightarrow e^+e^-$ reaction in M127-M118-Ker for $\nu_e$ ({\it Left}), $\bar\nu_e$ ({\it Middle}), and $\nu_x$ ({\it Right}, all species combined). The annihilation rate is computed by averaging energy deposition rates from 50 simulation snapshots $0.3\,{\rm \mu s}$ apart, $5\,{\rm ms}$ after merger, in a vertical slice of the simulation.}
\label{fig:pair_annihil}
\end{figure*}

Energy deposition into the polar region from $\nu\bar\nu\rightarrow e^+e^-$ pair annihilation can be an important effect in the production of low-density regions and relativistic outflows close to the rotation axis of the system~\cite{Fujibayashi:2017abc,Kawaguchi:2025con}. To evaluate the impact of pair annihilation in our simulations, we calculate the energy deposition from that reaction in M127-M118-Ker (using the opacity for pair annihilation computed from the isotropic kernel and form factor correction, as well as the energy density of neutrinos in the simulation). The results are shown on Fig.~\ref{fig:pair_annihil}. Pair annihilation is clearly active in the hot, dense regions of the remnant and in the hot spiral arms (especially for $\nu_e\bar\nu_e$), as well as in the polar regions within $\sim 10\,{\rm km}$ of the surface of the neutron star. At this very early time in the evolution of the post-merger remnant, however, the polar regions are still loaded with hot matter. For example, $20\,{\rm km}$ above the center of mass of the system, we have $\rho \sim 5 \times10^9\,{\rm g/cm^3}$, $T\sim 6.5\,{\rm MeV}$, and $Y_e\sim 0.35$. Under these conditions, the energy density of $\bar\nu_e$ in the simulation is about half of the energy density of neutrinos in statistical equilibrium with the fluid; the energy density of $\nu_e$ is close to its equilibrium value; the energy density of $\nu_x$ is about $3$ times smaller than its equilibrium value. The fluid is still hot enough that the $e^+e^-\rightarrow \nu\bar\nu$ reaction is faster than its inverse. The net effect of pair processes is thus to cool the outflows, albeit very slowly ($\sim$ tens of milliseconds, i.e. longer than it takes for the outflows to leave the part of the polar regions where these reactions matter). For electron-type neutrinos, the density also remains high enough that charged current reactions still dominate over pair processes. This is different from the situation observed at later times in~\cite{Fujibayashi:2017abc}, where pair annihilation was approximately included in a two-moment simulation, or~\cite{Kawaguchi:2025con}, which studied black hole-disk systems in axisymmetry with Monte Carlo transport. For example, in~\cite{Fujibayashi:2017abc}, pair annihilation was enough to accelerate the matter to mildly relativistic speeds, but the density of the fluid at the same point of the polar regions was nearly two orders of magnitude lower, and the temperature was $<2\,{\rm MeV}$.

We note that the energy deposition rate from pair annihilation in the polar region is roughly proportional to $L_\nu L_{\bar \nu}$, but will not change much as the density and temperature of the fluid in those regions decrease. Indeed, the distirbution function of neutrinos in those regions is set by the neutrinos emitted on the surface of the remnant and in the disk, not those emitted locally. By the time the temperature drops below $2\,{\rm MeV}$, neutrino emission becomes entirely negligible, while pair annihilation rates increase or decrease with $L_\nu L_{\bar \nu}$. At the order of magnitude level, the energy deposition rate observed in our simulation is consistent with what is seen $20\,{\rm ms}$ post-merger in~\cite{Fujibayashi:2017abc}, when $L_{\nu_e} L_{\bar \nu_e}$ is about the same as in our simulation (though with a higher $\nu_e$ luminosity and lower $\bar\nu_e$ luminosity). This is consistent with earlier results indicating comparing Monte Carlo and two-moment estimates of the energy deposition, indicating that differences of a factor of $2-3$ in energy deposition rates were the most likely outcome~\cite{Foucart:2018gis}. In this specific simulation, we would expect an increase in the energy deposition from pair annihilation over time (as our neutrino luminosity is increasing), up to the collapse of the remnant to a black hole.

\section{Conclusions}

In this manuscript, we present a new set of binary neutron star simulations using Monte Carlo neutrino transport, including the first merger simulation using the simulated energy distribution of neutrinos directly in the calculation of weak interaction rates. In that simulation, both inelastic scattering of neutrinos on electrons and $\nu\bar\nu \leftrightarrow e^+e^-$ interactions are calculated from that energy distribution, though with an approximate treatment of the dependence of those reactions on the angular distribution of neutrinos. Specifically, we assume an isotropic distribution of neutrinos for inelastic scattering, while for pair annihilation we use a single correction factor within each grid cell to account for the impact of the angular distribution of neutrinos. We additionally take advantage of a set of simulations slowly varying the total mass of the binary to assess the impact on the collapse time and the outflows of that parameter.

The use of the distribution function of neutrinos in weak reaction rates and, in particular, in blocking factors, is made possible by a reweighting of the neutrino packets in our Monte Carlo algorithm. Ideally, we would like each packet to represent a change $\Delta f_\nu \ll 1$ and $\Delta f_\nu \ll f_\nu$ in the distribution function $f_\nu$ of neutrinos. In practice, the best that can be done without greatly increasing the cost of simulations is to marginally satisfy $\Delta f_\nu < 1$ (after integration over the direction of propagation of the neutrinos). This allows us to use the $f_\nu$ measured in the simulation to calculate weak interaction rates in regions where neutrinos are out of equilibrium with the fluid, but not to directly resolve $f_\nu$ within a grid cell. Still, this is a significant improvement over previous simulations in which $\Delta f_\nu \gg 1$ was the norm and using $f_\nu$ in the calculation of blocking factors was hopeless~\cite{Foucart:2025nub}. There is however an important trade-off that has to be made to reach that objective: the shot noise in the coupling between neutrinos and the fluid (i.e. energy, momentum, and lepton number transfer) is larger with the reweighted packets.

The main effect of the inclusion of inelastic scattering and of our improved treatment of pairs is to increase the opacity of the fluid to heavy-lepton neutrinos, including in the densest regions of the hot tidal arms created after the merger. Inelastic scattering also leads to a $\sim 10\%$ decrease in the average energy and $\sim 30\%$ decrease in the total luminosity of escaping $\nu_x$ -- despite the remnant being hotter in that simulation as the neutron star approaches collapse. These changes in the interaction of heavy-lepton neutrinos with the fluid impact the ejection of matter in the immediate aftermath of the merger: we find $\sim 50\%$ more mass being ejected in the simulation with more advanced neutrino physics, though we caution that this relative difference will nearly certainly decrease as later-time outflows are taken into account. Overall, we thus find that the improved neutrino physics included in our simulation has more of an impact on the outflows that the use of Monte Carlo or two-moment schemes~\cite{Foucart:2024npn}. It does not change the qualitative results obtained in our simulation, and has less of an impact than e.g. the inclusion of magnetic fields (neglected in this manuscript); but it certainly appears relevant to any detailed quantitative modeling of merger outflows.

We also provide fits for the energy distribution of neutrinos in regions where they are out of equilibrium with the fluid. We find that heavy-lepton neutrinos have very non-thermal distributions in regions where scattering dominates over pair production processes. In very low-density regions, electron-type neutrinos are best modeled by relatively hard spectra, while heavy-lepton neutrinos have softer spectra. The energy deposition from pair annihilation in our simulations is broadly consistent with results obtained with a two-moment scheme~\cite{Fujibayashi:2017abc}, though at the relatively early times studied here it remains a subdominant effect on the evolution of polar regions that remain loaded with matter.

Finally, we note rapid variations in the properties of the matter outflows with the total mass of the system. The higher mass systems studied here produce mostly hot, high-$Y_e$ outflows ($Y_e\gtrsim 0.25$), with a subdominant contribution from cold, equatorial outflows. The lower mass system (total mass reduced by $\sim 10\%$) is instead dominated by $Y_e\sim 0.2$ outflows, with the high-$Y_e$ and low-$Y_e$ component offering subdominant contributions. The $Y_e\sim 0.2$ component is the most impacted by neutrino physics. Considering their geometry and energetics, the higher $Y_e$ component is likely produced during the violent oscillations of the neutron star remnant before its collapse to a black hole and the low-$Y_e$ component is due to the partial tidal distruption of the neutron stars. The intermediate component is broadly similar to the `spiral-wave-driven' outflows discussed in~\cite{Nedora:2019jhl,Radice:2023xxn}, though with a narrower $Y_e$ distribution in our simulation. We show that 2D visualizations of these outflows as a function of $Y_e, s, v, \cos\theta$ can help separating these components. We also note that our simulations have very similar average $Y_e$, but very different distributions in $Y_e$ that would certainly result in distinct nucleosynthetic yields. We thus emphasize that relying on average outflow properties in any modeling effort can miss important differences in the detailed properties of the outflows.

\acknowledgements

FF and SR acknowledge support from the Department of Energy, Office of Science, Office of Nuclear Physics, under contract number DE-SC0020435. FF acknowledges support from the National Science Foundation through grant 2510568. FF and MD acknowledge support from NASA through grant 80NSSC22K0719. P.C.-K.C. gratefully acknowledges support from NSF Grant PHY-2020275 (Network for Neutrinos, Nuclear Astrophysics, and Symmetries (N3AS)). MD acknowledges support from the National Science Foundation through Grant PHY-2407726.
L.K. acknowledges support from the National Science Foundation under Grants No. PHY-2407742; No. PHY-2207342;
and No. OAC-2513338 at Cornell. M.S. acknowledges
support from NSF grants PHY-2309211, PHY-2309231,
and OAC-2513339, and NASA award 80NSSC26K0340 at
Caltech. L.K and M.S. also thank the Sherman Fairchild
Foundation for their support.

\bibliography{References/References.bib}

\end{document}